\documentclass[longauth]{aa}
\usepackage{graphicx}
\usepackage{txfonts}

\usepackage{natbib}
\usepackage{hyperref} 
\usepackage{amsmath}
\hypersetup{colorlinks=true, linkcolor=blue, citecolor=blue, urlcolor=blue}

\usepackage{arydshln} % Required for dashed lines in tables

\usepackage{float}
\usepackage{amsmath}
\usepackage{amssymb}
\usepackage[colorinlistoftodos]{todonotes}
\usepackage{url}
\usepackage{comment}
\usepackage{array}
\usepackage{txfonts}
\usepackage{xcolor}
\usepackage{tabularx}
\usepackage{booktabs}
\usepackage{placeins}

\begin{document}

\title{Time resolved absorption of six chemical species with MAROON-X points to a strong drag in the ultra hot Jupiter TOI-1518\,b}

\author{A.~Simonnin
          \inst{1},
           V.~Parmentier
           \inst{1}, 
            J.P.~Wardenier
           \inst{2},
           G.~Chauvin
           \inst{1}, 
           A.~Chiavassa
           \inst{1}, 
           M.~N'Diaye
           \inst{1},
           X.~Tan
           \inst{3, 4},
           N.~Heidari
            \inst{5},
           B.~Prinoth
           \inst{6, 7}
           J.~Bean
           \inst{8},
            G.~H\'ebrard
            \inst{5,9},
            M.~Line
           \inst{10},
           D.~Kitzmann
           \inst{11}, 
           D.~Kasper
           \inst{8},
            S.~Pelletier
           \inst{12},
           J.V.~Seidel
           \inst{1,7},
           A.~Seifhart
           \inst{13},
            B.~Benneke
           \inst{2},
            X.~Bonfils
            \inst{14},
           M.~Brogi
           \inst{15, 16},
            J-M.~Désert
           \inst{17},
            S.~Gandhi
           \inst{18, 19}
           M.~Hammond
           \inst{20},
            E.K.H.~Lee
           \inst{11},
           C.~Moutou
            \inst{21},
           P.~Palma-Bifani
           \inst{1, 22},
            L.~Pino 
           \inst{23},
           E.~Rauscher
           \inst{24},
            M.~Weiner Mansfield
           \inst{8, 25},
           J.~Serrano Bell
            \inst{26}
           P.~Smith
           \inst{10},
          }

\institute{Laboratoire Lagrange,Université Côte d’Azur, Observatoire de la Côte d’Azur, CNRS, Nice, France\\
              \email{adrien.simonnin@oca. eu}
        \and
            Institut Trottier de Recherche sur les Exoplanètes, Université de Montréal, Montréal, Québec, H3T 1J4, Canada
        \and
            Tsung-Dao Lee Institute, Shanghai Jiao Tong University, 520 Shengrong Road, Shanghai, People’s Republic of China
        \and 
            School of Physics and Astronomy, Shanghai Jiao Tong University, 800 Dongchuan Road, Shanghai, People’s Republic of China
        \and
            Institut d'astrophysique de Paris, UMR7095 CNRS, 
            Universit\'e Pierre \& Marie Curie, 
            98bis boulevard Arago, 75014 Paris, France 
        \and
            Lund Observatory, Department of Astronomy and Theoretical Physics, Department of Physics, Lund University, Lund, Sweden
        \and 
            European Southern Observatory, Alonso de Córdova 3107, Vitacura, Región Metropolitana, Chile
        \and
            Department of Astronomy \& Astrophysics, University of Chicago, Chicago, IL 60637, USA
        \and
        %2
            Observatoire de Haute-Provence, CNRS, Universit\'e d'Aix-Marseille, 04870 Saint-Michel-l'Observatoire, France
        \and 
            School of Earth and Space Exploration, Arizona State University, Tempe, AZ 85281, USA
        \and
            Center for Space and Habitability, University of Bern, Gesellschaftsstrasse 6, CH-3012 Bern, Switzerland
        \and
            Observatoire astronomique de l’Université de Genève, 51 chemin Pegasi 1290 Versoix, Switzerland
        \and
            Gemini Observatory/NSF NOIRLab, 670 N. A'ohoku Place, Hilo, HI 96720, USA
        \and
        %3
            Univ. Grenoble Alpes, CNRS, IPAG, 414 rue de la Piscine, 38400 St-Martin d'H\`eres, France 
        \and
            Department of Physics, University of Turin, Via Pietro Giuria 1, I-10125, Turin, Italy
        \and
            INAF – Osservatorio Astrofisico di Torino, Via Osservatorio 20, I-10025, Pino Torinese, Italy
        \and 
            Anton Pannekoek Institute of Astronomy, University of Amsterdam, Amsterdam, Netherlands
        \and 
            Department of Physics, University of Warwick, Coventry CV4 7AL, UK
        \and     
            Centre for Exoplanets and Habitability, University of Warwick, Gibbet Hill Road, Coventry CV4 7AL, UK
        \and 
            Atmospheric, Oceanic, and Planetary Physics, Department of Physics, University of Oxford, Parks Rd, Oxford OX1 3PU, UK
        \and
        %4
            Universit\'e de Toulouse, CNRS, IRAP, 14 avenue Belin, 31400 Toulouse, France
        \and 
            LESIA, Observatoire de Paris, Univ PSL, CNRS, Sorbonne Univ, Univ de Paris, 5 place Jules Janssen, 92195 Meudon, France
        \and
            INAF-Osservatorio Astrofisico di Arcetri Largo Enrico Fermi, Florence, Italy
        \and 
            Department of Astronomy, University of Michigan, Ann Arbor, MI 48109, USA
        \and 
            Steward Observatory, University of Arizona, Tucson, AZ, USA
        \and
            International Center for Advanced Studies (ICAS) and ICIFI (CONICET), ECyT-UNSAM, Campus 
            Miguelete, 25 de Mayo y Francia, (1650) Buenos Aires, Argentina              
}    
 
\titlerunning{Time resolved absorption in TOI-1518b}
\authorrunning{A. Simonnin et al.}
\date{Received November 30, 2024}

 \abstract

   {Wind dynamics play a pivotal role in governing transport processes within planetary atmospheres, influencing atmospheric chemistry, cloud formation, and the overall energy budget. Understanding the strength and patterns of winds is crucial for comprehensive insights into the physics of ultra-hot Jupiter atmospheres. Current research has proposed different mechanisms that limit wind speeds in these atmospheres.
}
   {This study focuses on unraveling the wind dynamics and the chemical composition in the atmosphere of the ultra-hot Jupiter TOI-1518\,b.}
   {Two transit observations using the high-resolution ($R_{\lambda}\sim85$\,000), optical (spectral coverage between 490 and 920\,nm) spectrograph MAROON-X were obtained and analyzed to explore the chemical composition and wind dynamics using the cross-correlation techniques, global circulation models (GCMs), and atmospheric retrieval. }
   {We report the detection of 14 species in the atmosphere of TOI-1518\,b through cross-correlation analysis. { Vanadium Oxide was detected only with the new HyVO line list whereas Titanium Oxide was not detected.} Additionally, we measure the time-varying cross-correlation trails for 6 different species, compare them with predictions from GCMs and conclude that a strong drag is slowing the winds in TOI-1518\,b's atmosphere ($\tau_{\rm drag}\approx 10^3-10^4s$). { We find that the trails are species dependent. Fe+ favors stronger drag than Fe, which we interpret as a sign of magnetic effects being responsible for the observed strong drag. Furthermore, we show that Ca+ probes layers above the Roche lobe, leading to a qualitatively different trail than the other species. Finally, We use a retrieval analysis to further characterize the abundances of the different species detected. Our analysis is refined thanks to the updated planetary mass of $1.83 \pm 0.47$~M$_{\rm{Jup}}$ we derived from new Sophie radial-velocity observations.  We  {measure} an abundance of iron of $ {\log_{10}{Fe}}=-4.88^{+0.63}_{-0.76}$ corresponding to 0.07 to 1.62 solar enrichment. For the other elements, the retrievals appear to be biased, probably due to the different K$_{\rm{p}}$/V$_{\rm{sys}}$ shifts between iron and the other elements, which we demonstrate for the case of VO.} 
}
   {}

\maketitle
%
%-------------------------------------------------------------------

\section{Introduction}

The golden age of exoplanet characterisation began in the last two decades. One of the most exciting topics is the exploration of their atmospheric diversity in terms of composition and dynamics
\citep{2019ARA&A..57..617M,2022ARA&A..60..159W}. Exoplanet atmosheres can be studied by observing their spectra either in emission \citep{2005A&A...438L..29C,2008ApJ...674..482S}, in transmission \citep{2002ApJ...568..377C} and soon in reflected light \citep{2013MNRAS.436.1215M} from ground-based and space observatories. Recently, the sensitivity of the JWST has begun to revolutionize this field, enabling extremely advanced studies of the fine structure and dynamics of the atmospheres of giant planets \citep{2023Natur.617..483T,2023Natur.620..292C}. Thanks to their extended atmospheres, ultra-hot Jupiters (UHJs) are ideal targets for atmospheric characterization in transmission. These planets are very close to their stars and are tidally locked. This implies a significant day/night temperature gradient, which creates strong atmospheric circulation \citep{2020SSRv..216..139S}. Differences in temperature of several hundred degrees have been measured between the daysides and nightsides \citep{2018haex.bookE.116P}. 
Due to their extreme temperature, volatile and refractory elements are accessible and detectable in such atmospheres. Indeed refractory species (with high condensation temperature) are expected to be gaseous in UHJs \citep{2018ApJ...866...27L} when, in colder planets they are inaccessible because they condensed out of the gas phase. The measure of the refractory to volatile elemental ratio of these planets (e.g., O/Fe. C/Fe), recently emerged as a new powerful way to trace planet formation \citep{2021ApJ...914...12L,2023ApJ...943..112C,2025AJ....169...10P,2025AAS...24511904S}.
When using low to moderate spectral resolution ($R_{\lambda}<5\,000$), (e.g. JWST), the observed spectra contains a mixture of information from various parts of the atmosphere. Because each part of the atmosphere has different properties, such as temperature or chemical composition \citep{2024Natur.632.1017E}, this can lead to misleading or biased inferences about the different properties derived from the data \citep{2015ApJ...800...22F,2016ApJ...820...78L}.\\
For the first time in 2010, high-resolution ($R_{\lambda}>40\,000$) spectroscopy was used to characterize the atmosphere of a transiting hot Jupiter by resolving individual molecular lines using CRIRES at VLT \citep{2010Natur.465.1049S}. During the transit, the Doppler shifts caused by the planet's rotation and the atmospheric winds allow lines formed in different parts of the planetary atmosphere to be spectroscopically separated (e.g.  \citealt{2025A&A...693A.213N}). Recent ESPRESSO observations at VLT showed that the iron absorption lines of WASP-76b and WASP-121b, two canonical Ultra-Hot Jupiters (UHJs), are progressively blueshifted during the transit \citep{2020Natur.580..597E,2021A&A...645A..24B}.
While different scenarios have been suggested to explain this behavior, the precise physical mechanism remains elusive.  
The signal could result from a hot, puffy evening terminator and a cool, compact morning terminator, whereby the blueshifting winds and rotation of the hot evening terminator dominate the absorption signal due to its larger-scale height \citep{2021MNRAS.506.1258W}. Alternatively, it has been shown that 3D models with opaque clouds can also reproduce the observed signal \citep{2022ApJ...926...85S}, whereby the cloud deck “blocks” the absorption features on the morning terminator. Other studies highlights the richness of WASP-76b with the detection of multiple species with different shift \citep{2022AJ....163..107K}. {Finally, as shown by \citet{2023AJ....165..257B}, magnetic effects may also contribute to the observed Doppler shift in high-temperature targets.}\\

Here we present two transit observations of TOI-1518\,b with MAROON-X at the Gemini-North Observatory. As show in Table\,\ref{Sysparam}, TOI-1518\,b, with an equilibrium temperature of (T$_\textrm{eq}$ = 2546 K) sits in-between the well studied UHJs WASP-76\,b (T$_\textrm{eq}$ = 2228 K) and WASP-121\,b (T$_\textrm{eq}$ = 2720 K) with iron previously detected by \citet{2021AJ....162..218C}.

After presenting the observations and the data reduction in Sect.\,2, we present the chemical information we obtained thanks to Cross-Correlation techniques in Sect.\,3. Then, we compare the iron trail detected with global circulation models (GCMs) to explore the wind dynamics of the planet in Sect.\,4. We finally present a retrieval analysis in Sect.\,5, hinting at the different abundances of the species detected in the atmosphere of TOI-1518\,b.

\section{Observations and data reduction}

We observed two transits of the UHJ TOI-1518\,b with MAROON-X, a high-resolution ($R_{\lambda} \sim 85\,000$) optical (spectral coverage between 490 and 920\,nm) spectrograph at the 8.1-m Gemini-North observatory in Hawaii. Recent observations showed the capacity of MAROON-X to characterize UHJs by detecting ions and volatile and refractory elements. It has also allowed the description of the time variation of the atmospheric signal during transit (also known as the "trail") \citep{2023Natur.619..491P,2023A&A...678A.182P} and even to study some strong lines such as Ca+ triplet \citep{2024A&A...685A..60P}.
The observations were taken on 2022-08-13 and 2023-10-19 (program ID GN-2022B-Q-128 and GN-2023B-Q-127, PI: Parmentier). A summary of the observations is given in Table\,\ref{KapSou}.

%--------------------------------------------------- One column table

\begin{figure*}[!thbp]
   \centering
   \includegraphics[width=6cm]{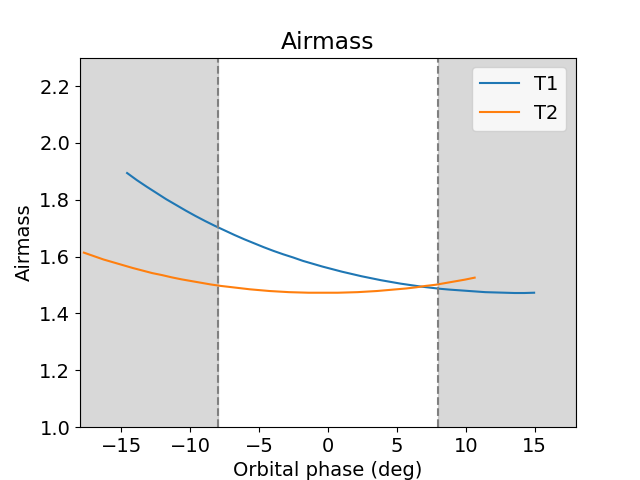}
   \includegraphics[width=6cm]{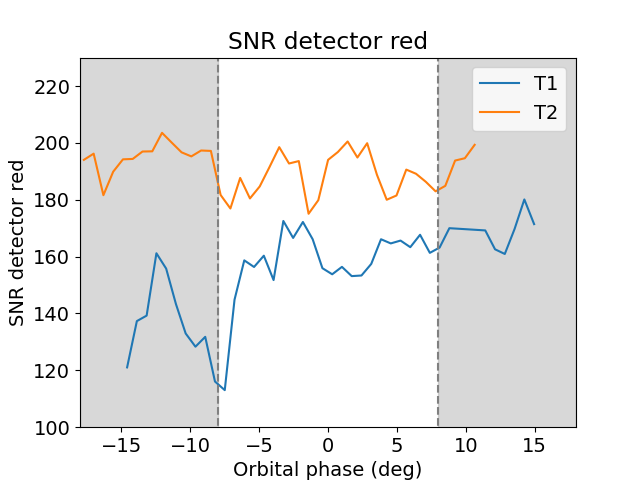}
   \includegraphics[width=6cm]{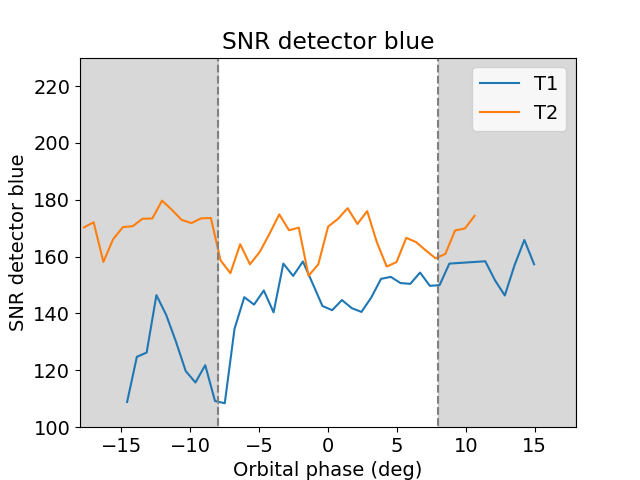}
   \caption{Airmass and S/N plot over orbital phases. The gray area represents the out-of-transit phases. Due to better observational conditions (humidity and airmass), S/N is higher for the second transit.}
   \label{FigAirmass}%
    \end{figure*}

\begin{table}[!htbp]
    \centering
    \caption[]{Overview of TOI-1518\,b observations during the 2 transits from Programme ID GN-2022B-Q-128 and GN-2023B-Q-127, PI: Parmentier}
    \label{KapSou}
    \begin{tabular}{c c c}
        %\hline
        \toprule
        \noalign{\smallskip}
        Night      &  2022-08-13  &  2023-10-19  \\
        \noalign{\smallskip}
        Phase  &  0.96-0.04   &  0.96-0.04 \\
        \noalign{\smallskip}
        N$_{obs}$  &  40 (25 + 15) &  41 (25 + 16) \\
        \noalign{\smallskip}
        Exp. time &  260s (b), 220s (r)&  220s (both arms)  \\
        \noalign{\smallskip}
        Airmass   &  1.45 - 1.9 &  1.47 - 1.6  \\
        \noalign{\smallskip}
        S/N  &  110 - 180  &  155 - 210  \\
        & (Avg = 150) & (Avg = 180)\\
        \noalign{\smallskip}
        \bottomrule
        %\hline
        \noalign{\smallskip}
    \end{tabular} \\ 
\textbf{Notes.} N$_{obs}$ is the total number of observed spectra with in-transit (25) and out-of-transit (15) observations
\end{table}
MAROON-X is divided into two detectors, one "blue" covering wavelengths ranging from 490 to 678 nm. The second one, "red," covers wavelengths ranging from 640 to 920 nm. To ensure complete coverage of transit events, the observations include 30min baseline measurements both pre- and post-transit.
The exposure time for the red detector was slightly lower than for the blue arm (220s vs 260s) during the first observation. For the second observation, the exposure time for both detectors was set up to 220 sec. Each order of the red detector comprises 4036 pixels, while the blue one has only 3954 pixels. There are 28 spectral orders for the red detector and 33 for the blue one. Fig \ref{FigAirmass} presents the signal-to-noise ratio (S/N) and the airmass as a function of the observation frames for both transits. For both nights, the S/N was was always above 110, with a slightly better S/N for the second night (155 min and 210 max).  
%This could be due to the better airmass conditions of the second transit.
During the second night, the blue arm looked less performant than the red arm because the blue arm's exposure time was higher during the first transit. Conditions of the second transit (average humidity = 7\% and lower airmass) were better and so both arms still have better S/N than during the first transit (average humidity =24\%). 

\begin{figure*}[!thbp]
   \centering
   \includegraphics[trim = 0.25cm 0.25cm 0.25cm 0.25cm, clip,width=18cm]{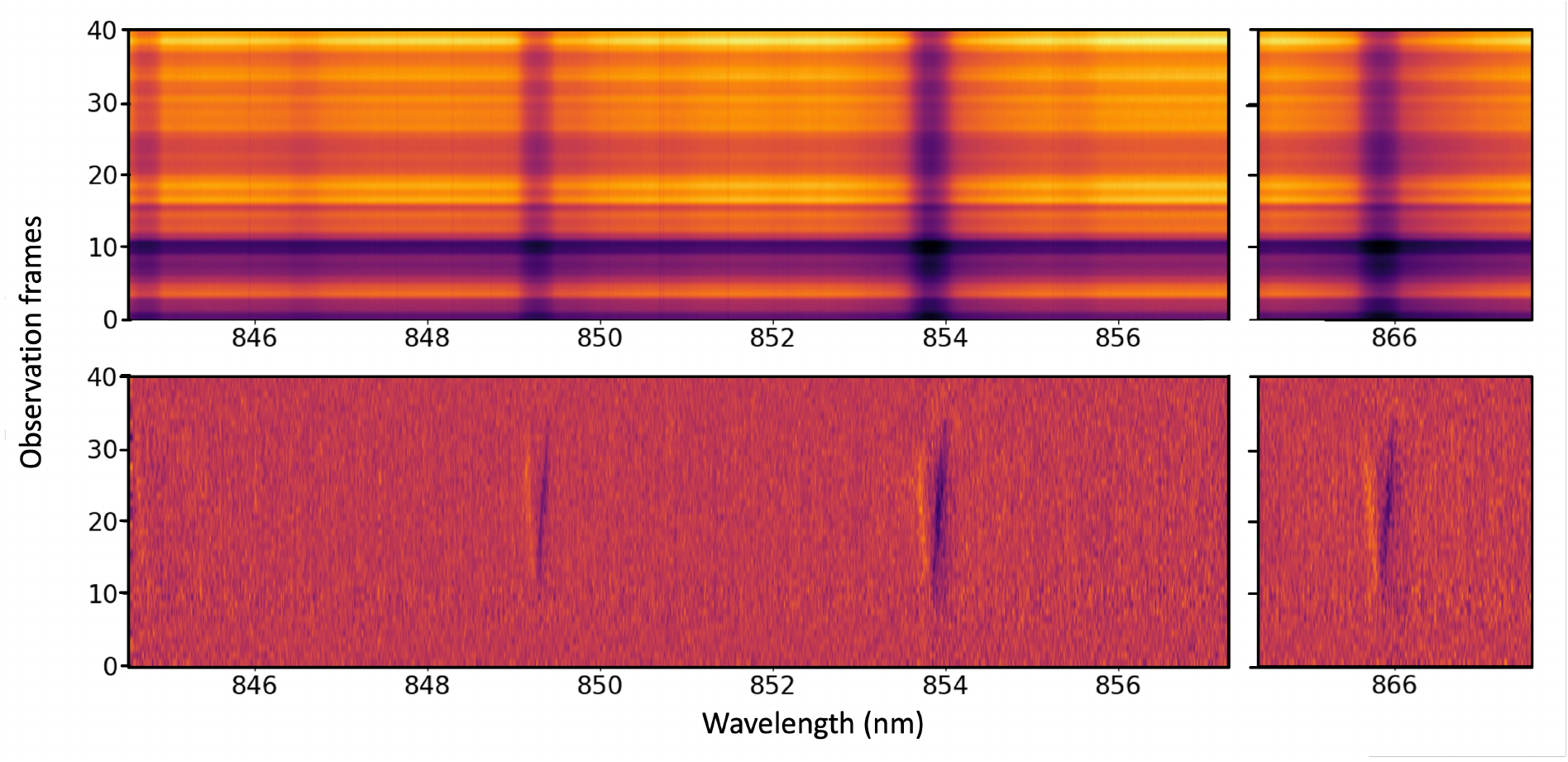}
   \caption{Correction of the stellar and telluric signal. {\bf Top Panel}: Raw data of two orders of MAROON-X red detector between 845 to 867 nm. Three strong stellar lines corresponding to Ca+ triplet are observed around 849, 854 and 866 nm. {\bf Bottom panel}: Residuals obtained after correction with out-of-transit data. The dark signal here is the planetary Ca+ absorption lines, while the yellow signal is the Doppler shadow discussed in Sect. \ref{result_trail} due to the Rossiter-McLaughlin effect.}
    \label{FigCa+}%
\end{figure*}

\color{blue}

\color{black}

The MAROON-X data were reduced using the standard pipeline \citep{2020SPIE11447E..1FS} in one-dimensional wavelength-calibrated spectra, order by order for each time series exposure. The outputs are given as $\textrm{N}_{\textrm{orders}} \times \textrm{N}_{\textrm{frames}} \times \textrm{N}_{\textrm{pixels}}$ with $\textrm{N}_{\textrm{orders}}$ the number of spectral orders.
In total, 40 frames were observed during both transits. The redder order of the blue detector (between 668 and 678nm) has been removed because of a too-low S/N (< 35). 

One main limitation with high-resolution transmission spectroscopy from ground-based observations is the Earth’s atmosphere and stellar signals. Planetary signals are much fainter but change over time because of the rapid Doppler acceleration, inducing shifts of many tens of km.s$^{-1}$ to the planet spectrum over the transit duration. The telluric lines, however, stay constant, and the positions of the stellar lines vary only on the order of hundreds of m.s$^{-1}$. This allows us to distinguish the planetary signal from the other two. We applied different reduction steps on the data following the sequence described by \citet{2023Natur.619..491P} and summarized below: %(See for more details) 
\begin{itemize}
\item All observed spectra are aligned in the stellar rest frame to remove the Earth’s barycentric motion and TOI-1518’s reflex motion. This is necessary to subtract the stellar signal from the data. 
\item Each spectrum is set to the same continuum level to remove blaze and throughput variations. 
\item The in-transit data are divided by a master stellar spectrum made with the averaged out-of-transit data.
\item A principal component analysis (PCA) approach removes the telluric signal and residuals from the stellar correction. { For most species we removed the first three principal components and verified that this was not significantly affecting our signal (see Fig. \ref{Diff_comp}). This is not the case for the Ca+ lines, which dominate the observed spectrum and are thus too affected by the PCA. For these we decided to not use any PCA correction. }
\end{itemize}

{  We show in Fig.\,\ref{FigCa+} two orders of MAROON-X red detectors before (top) and after (bottom) the division by the master out stellar spectrum. The strong stellar Ca+ lines are efficiently removed, leaving apparent the planetary lines and the Doppler shadow effect (see sub-Section \ref{result_trail}). In this wavelength range, the planetary Ca+ lines are clearly visible, even without PCA corrections.}

\section{Cross-correlation analysis}

\subsection{TOI-1518 system}
\begin{table}[!th]
\begin{center}
  \caption[]{TOI-1518 stellar and planetary parameters from \cite{2021AJ....162..218C}.
  {Recent observations from SOPHIE (describe in Sect. \ref{Sophie}) provide updates for some of them (marked with $^a$).}}
  \label{Sysparam}
  \begin{tabularx}{250pt}{X X}

    \toprule
    \noalign{\smallskip}
    Stellar parameters & Value  \\
    \noalign{\smallskip}

    \midrule
    \noalign{\smallskip}
    Stellar radius & $1.950 \pm 0.048\,R_\odot$  \\
    \noalign{\smallskip}
    Effective temperature & $7300 \pm 100\,\text{K}$   \\
    \noalign{\smallskip}
    Metallicity [Fe/H] & $-0.1 \pm 0.12\,\text{dex}$   \\
    \noalign{\smallskip}
    Rotational velocity & $85.1 \pm 6.3\,\text{km.s}^{-1}$   \\
    \noalign{\smallskip}
    Spectral type & F0   \\
    \noalign{\smallskip}

    \midrule
    \noalign{\smallskip}
    Planetary parameters & Value  \\
    \noalign{\smallskip}

    \midrule
    \noalign{\smallskip}
    Planet mass & $1.83 \pm 0.47\,M_\text{J}^{a}$  \\
    \noalign{\smallskip}
    Planet radius & $1.875 \pm 0.053\,R_\text{J}$   \\
    \noalign{\smallskip}
    Equilibrium temperature & $2546^{+35}_{-36}\,\text{K}^{a}$    \\
    \noalign{\smallskip}
    \midrule
    \noalign{\smallskip}
    System parameters & Value  \\
    \noalign{\smallskip}
    \midrule
    \noalign{\smallskip}
    Orbital period & $1.90261131 \pm 0.00000043\,\text{days}^{a}$  \\
    \noalign{\smallskip}
    Mid-transit Time & $2458787.04943 \pm 0.00028\,\text{BJD\_TDB}^{a}$   \\
    \noalign{\smallskip}
    Orbital Inclination & $77.626\pm 0.097\,\text{degrees}^{a}$    \\
    \noalign{\smallskip}
    Semi-major axis & $0.03712 \pm 0.00082\,\text{AU}^{a}$   \\
    \noalign{\smallskip}
    Systemic Velocity ($V_\text{sys}$) & $-13.94 \pm 0.17\,\text{km.s}^{-1}$   \\
    \noalign{\smallskip}
    Projected orbital velocity ($K_\text{p}$) & 
    $207.32 \pm 4.51\,\text{km.s}^{-1}$  $\star$  \\ 
    \noalign{\smallskip}
    Impact Parameter (b) & $0.9036^{+0.0061}_{-0.0053}$   \\
    \noalign{\smallskip}
    \bottomrule
    \bottomrule
  \end{tabularx}
  \end{center}
  {\textbf{Notes:} $\star$ : derived from Eq.\ref{eu_eqn}.}
\end{table}

{TOI-1518\,b is an ultra-hot Jupiter that was discovered by \citet{2021AJ....162..218C}. It is a 1.875 R$_{\textrm{Jup}}$ planet with an equilibrium temperature for zero albedo of 2546 K. The planet is misaligned, with an impact parameter of 0.9 (see Table.\ref{Sysparam}), meaning that it doesn't cross the zero-velocity  {plane} of the star. The planet is precessing and the impact parameter has been shown to vary through time \citep{2024PASJ..tmp...34W}. Between our observations a change of  {0.03} was expected, which is too small to affect our observations(from b = 0.91497 in 2019 to b= 0.8797 in 2022).The geometry of the system is presented in Fig.\,\ref{FigGeometry}.}  

{The planet orbit a 7300\,K F-type star, which is fast rotating. The fast rotation rate from the star impeded \citet{2021AJ....162..218C} to detect the radial velocity motion due to the orbit of the planet. As such, \citet{2021AJ....162..218C} were only able to put an upper limit to the planetary mass.
We also report in Appendix \ref{Sophie} the first significant detection of TOI-1518\,b in radial velocity, using the SOPHIE spectrograph. This allows us to even more validate the planetary nature of the transits and to better characterize the system's parameters, in particular the planetary mass and the semi-major axis.
SOPHIE measurements allow us to pin-down the mass of TOI-1518\,b to 1.83 $\pm 0.47$ M$_{\textrm{Jup}}$ (see Appendix \ref{Sophie} for details). Additionally, we analyzed 56 transits of TOI-1518\,b captured by TESS, compared to only 24 analyzed in \citet{2021AJ....162..218C}. This increased data set has enabled us to refine the planetary ephemerides, including the semi-major axis. The improved parameters lead to a planetary semi-amplitude velocity Kp=207.32$\pm 4.51$, which is 10 km/s lower (1.5 sigma lower) than in \citet{2021AJ....162..218C}}

\subsection{Template spectra for cross-correlation} \label{sectionCCF} 
\begin{figure*}[!thbp]
\centering
  \includegraphics[trim = 1.5cm 0.25cm 1cm 2.3cm, clip,width=18.7cm]{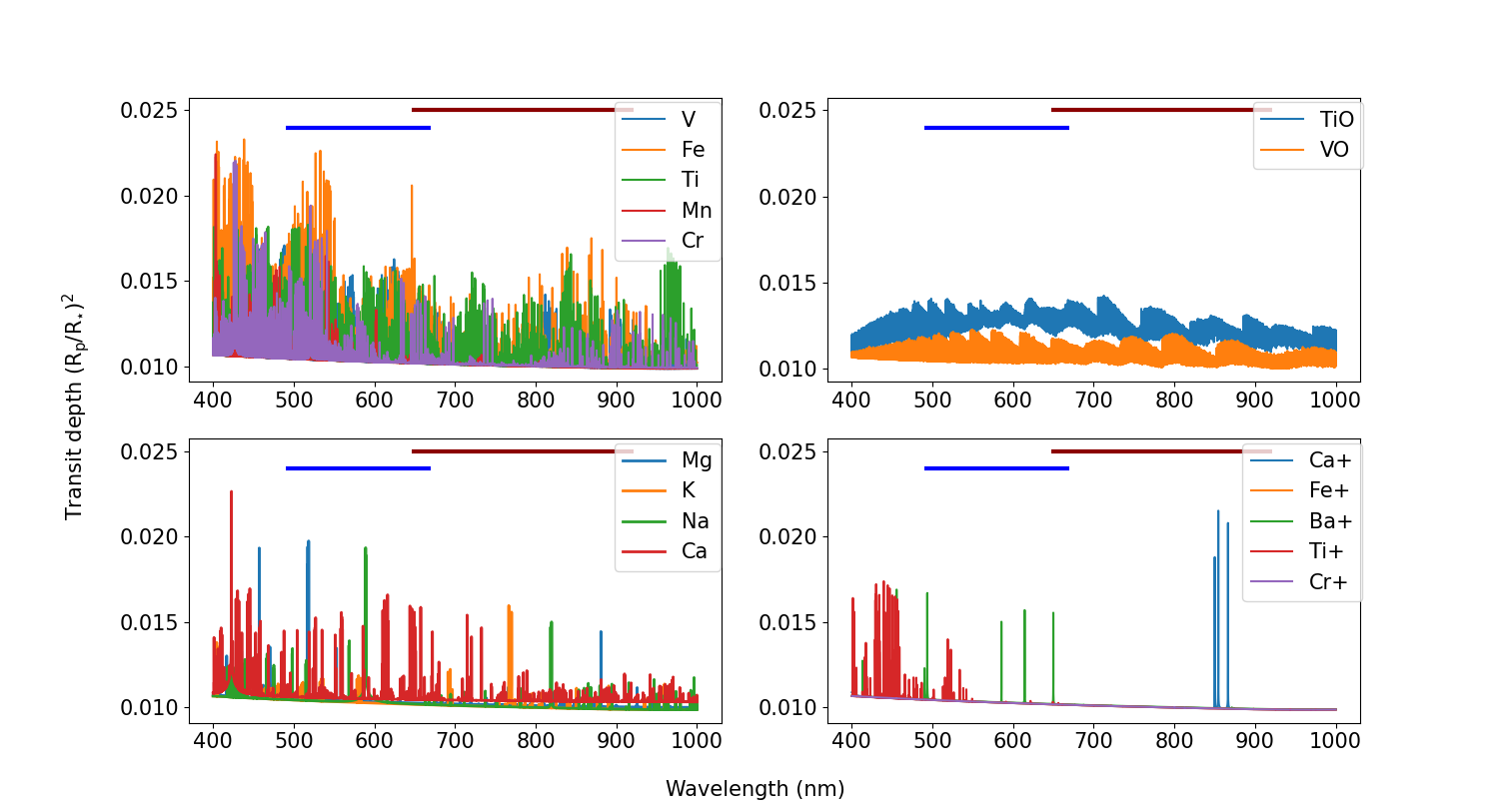}
  \caption{Synthetic transmission spectra computed with PetitRadtrans and FastChem used as template for the cross-correlation anaysis. They are computed for TOI-1518\,b parameters, assuming a temperature of 2500 K. The wavelength coverage of MAROON-X used in this study is represented with the blue (blue detector, 490-670 nm) and the red line (red detector, 640-920 nm). Ions and alkalines, have few very strong lines, while other metals are composed of line forests. Molecules also have forests of spectral lines grouped in distinct absorption bands. Except for the Ca+ lines, the other metals present strong signals in the range of the blue detector of MAROON-X, where few telluric lines are present. At 2500K, the Fe+ and $Cr+$ lines are smaller than the others and are not visible in this plot. }
     \label{Fig_PetitRadtrans}
\end{figure*}

The Cross-Correlation method is needed to detect the faint planetary lines in the residuals obtained after PCA. We cannot detect most of them directly except for a few individual lines (such as Ca+). Fortunately, atoms and molecules have many spectral lines and a cross-correlation with a template boosts the signal, allowing a precise detection of the planetary spectrum. A template is needed to combine these lines. We generated synthetic spectra specific to TOI-1518\,b using the parameters from Table.\,\ref{Sysparam} and PetitRadtrans \citep{2019A&A...627A..67M}, assuming a temperature of 2500K, with chemical abundances determined by equilibrium chemistry and calculated with FastChem \citep{2022MNRAS.517.4070S}.
We produced spectra for individual molecules at a resolution of $R_{\lambda} = 250\,000$ over the 400 to 1000\,nm wavelength range. These single-species spectra, to be used as cross-correlation templates, use collision induced absorption (CIA) cross-sections of H2-H2 and H2-He as continuum. We then interpolated the spectra onto the MAROON-X wavelength grid and convolved them to match the instrumental resolution.
The spectra were also convolved with the planetary rotation kernel. These spectra are shown in Fig.\,\ref{Fig_PetitRadtrans}. The species selected in this study are based on those previously detected in recent MAROON-X publications \citep{2023A&A...678A.182P,2023Natur.619..491P}. \\ Line list from Kurucz database were used for all atoms and ions \citep{2017CaJPh..95..825K}. For TiO, we used TOTO line list \citep{2019MNRAS.488.2836M}.  For VO, both the  HyVO line list \citep{2024MNRAS.529.1321B} and the VOmyt line list \citep{2016MNRAS.463..771M} outputs were compared in Sect.\ref{KpVsys} and \ref{disc_CCF}.
\\
Alkaline metals and ions show individual strong lines, while other metals show line forests. Most of the signals except for the Ca+ triplet are stronger in the blue part of the spectrum, which corresponds to the blue detector of MAROON-X. We decided to analyze each detector individually as if it were two different transits and then to sum every $K_p$-$V_{\rm res}$ map or Cross-Correlation Function (CCF) map.

\subsection{Species detected in the $K_p$-$V_{\rm res}$ maps}  \label{KpVsys}
\begin{figure*}[!thbp]
   \centering
      \includegraphics[width=18cm]{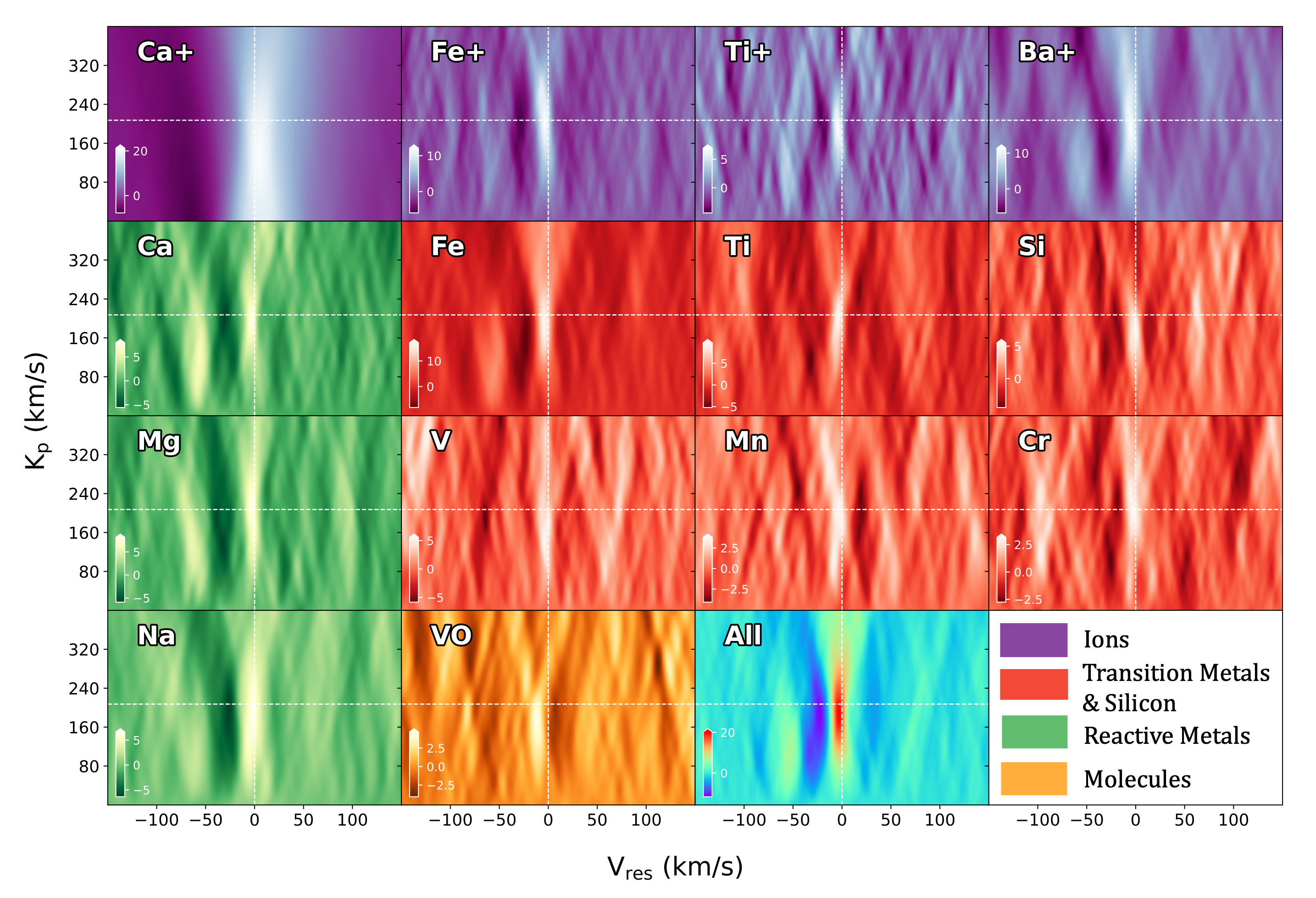}
      \caption{\textbf{All K$_{p}-V_{res}$ diagram for detected species in TOI-1518\,b dataset}. The white cross indicates the expected location of the planetary signal, which assumes a static atmosphere. Deviations from the white cross could be the significance of wind, circulations, or chemical asymmetries on TOI-1518\,b. A clear signal is observed with a white blob, sometimes shifted, near the white cross in each diagram.  
      The signal observed at $K_{p}$ around 100 km.s$^{-1}$ and $V_{\rm res}$ around -60 km.s$^{-1}$ in some diagrams is due to the Doppler shadow and is not of planetary origin. Note that the Ca+ K$_{p}-V_{res}$ map was computed without the use of PCA (see discussion). 
              }
         \label{All_Kp_Vrest}
\end{figure*}

The cross-correlation maps are converted into velocity-velocity maps ($K_p$-$V_{\rm res}$ diagrams) by shifting them to the expected rest-frame of the planet {($K_p$ = 207.32 $\pm$ 4.51\,km.s$^{-1}$, $V_{\rm sys}$ = $-13.94$ $\pm$ 0.17\,km.s$^{-1}$)}, assuming values of projected orbital velocity between 0 and 400\,km.s$^{-1}$ in steps of 1 km. s$^{-1}$. Fig.\,\ref{All_Kp_Vrest} shows fourteen clear detections obtained via cross-correlating the signal with single-species templates. The white cross in each plot is the position of the expected signal from the planet if the planetary atmosphere is considered static and the planet has a circular orbit. The expected $K_p$ is calculated as follows:  
{\begin{equation} \label{eu_eqn}
K_{p} = V_{orb} * \rm{sin}(i) = 212.25 * \rm{sin}(77.626) = 207.32 \pm 4.51\,km.s^{-1}
\end{equation} }
with $V_{orb}=\frac{2\pi a_{m}}{P}$, where $P$ is the period, $a_{m}$ is the semi-major axis of TOI-1518\,b and $i$ the inclination of the system. All these parameters are given in Table\, \ref{Sysparam}. To compute our $K_p$-$V_{\rm res}$ plots, we selected a range of orbital velocity ($K_p$) from 0 to 400 km.s$^{-1}$ and a range of rest-frame velocity ($V_{\rm res}$) from -150 to 150\, km.s$^{-1}$  with steps of 1 km.s$^{-1}$ for each. We then integrate each point of the CCF maps previously obtained (see Fig.\,\ref{All_Kp_Vrest}) following the slope determined by the orbital velocity at the rest frame position determined by $V_{\rm res}$. The noise level is calculated in a region far from the central peak at $V_{\rm res}$ $\ge$ 75 km.s$^{-1}$ where no signal of the planet or Rossiter-McLaughlin (RM) residuals is expected. Fourteen species are detected with an S/N $\ge$ 4. 
The parameters of the best Gaussian fits of the $K_p-V_{\rm res}$ maps are presented in Table\,\ref{tab_KP_Vsys}.

\begin{table}[!htbp]
    \begin{center}
    \caption[]{Best fit parameters of the Gaussian fits to the $K_p$-$V_{\textrm{res}}$ diagram of Fig.~\ref{All_Kp_Vrest} at the $K_p$ position where the maximum signal is observed. }
    \begin{tabular}{c c c c c c}
        %\hline
        \toprule
        \noalign{\smallskip}
         & $\Delta$ K$_p$ &  Amp ($\sigma$)  &  V$_{\textrm{res}}$ & FWHM  \\
        %\hline
        \midrule
        \noalign{\smallskip}
        Ca+  &  -74.1 $\pm$ 0.5 & 21.1 $\pm$ 0.5 & 4.9 $\pm$ 0.1 & 36.9 $\pm$ 0.2\\
        \noalign{\smallskip}
        Fe+  &  -7.1 $\pm$ 0.1 &  12.3 $\pm$ 1.1   & -4.0 $\pm$ 0.5 & 11.4 $\pm$ 1.2\\
        \noalign{\smallskip}
        Ti+ &  -16.0 $\pm$ 0.1 &  7.5  $\pm$ 1.1  & -3.4 $\pm$ 0.6 & 9.1 $\pm$ 1.4\\  
        \noalign{\smallskip}
        Ba+ &  -15.9 $\pm$ 0.1  &  11.4  $\pm$  -0.6 & -5.2 $\pm$ 0.3 & 13.7 $\pm$ 0.8\\
        \noalign{\smallskip}
        Ca &  -11.4 $\pm$ 0.1  &  8.1 $\pm$ 0.9  & -3.7 $\pm$ 0.6 & 10.1 $\pm$ 1.4\\
        \noalign{\smallskip}
        Fe  & -22.2 $\pm$ 0.2  &  17.9 $\pm$ 1.7  & -3.8 $\pm$ 0.5 & 11.1 $\pm$ 1.2\\
        \noalign{\smallskip}
        Ti  &  -4.5 $\pm$ 0.4  &  10.1 $\pm$ 0.7 &-4.3 $\pm$ 0.3 & 9.7 $\pm$ 0.8\\
        \noalign{\smallskip}
        Si  &  -29.2 $\pm$ 0.2  &  5.9  $\pm$ 0.9  & -1.4 $\pm$ 0.8 & 9.7 $\pm$ 1.8\\
        \noalign{\smallskip}
        Mg  &  -6.5 $\pm$ 0.3  &  8.1 $\pm$ 0.8 & -3.1 $\pm$ 0.6 & 11.7 $\pm$ 1.4\\
        \noalign{\smallskip}
        V  &  -47.4 $\pm$ 0.6 &  5.8 $\pm$ 0.5  & -3.1 $\pm$ 0.5 & 10.2 $\pm$ 1.1\\
        \noalign{\smallskip}
        Mn  &  -5.5 $\pm$ 0.6  &  4.0 $\pm$ 0.2 & -3.5 $\pm$ 0.2 & 12.7 $\pm$ 0.6\\
        \noalign{\smallskip}
        Cr  &  -8.8 $\pm$ 0.1  &  4.0 $\pm$ 0.4 & -1.3 $\pm$ 0.8 & 12.5 $\pm$ 1.9\\
        \noalign{\smallskip}
        Na  &  -14.2 $\pm$ 0.1 &  7.9 $\pm$ 1.0  & -1.3 $\pm$ 1.2 & 18.3 $\pm$ 3.3\\
        \noalign{\smallskip}
        VO  &  -32.8 $\pm$ 0.2 &  4.9 $\pm$ 0.2  &-12.1 $\pm$ 0.2 & 10.5 $\pm$ 0.5\\
        \noalign{\smallskip}
        All  &  -20.5 $\pm$ 0.1 &  21.1 $\pm$ 2.6  & -2.9 $\pm$ 0.7 & 11.6 $\pm$ 1.7\\
        \noalign{\smallskip}
        %\hline
        \bottomrule
        \noalign{\smallskip}
    \end{tabular} \\ 
    \label{tab_KP_Vsys}
    \end{center}
    {\textbf{Notes :} Amplitude (Amp) corresponds to the best-fit line depth of the absorbing species above the spectral continuum. The amplitude is expressed in units of signal-to-noise ratio ($\sigma$), as each map is divided by the standard deviation as explained in Sect. \ref{sectionCCF}. Detections range from 4 to 21.1 $\sigma$. \color{black}  $V_{\textrm{res}}$ corresponds to the radial velocity of the line center, as measured in the rest frame of the stellar system (same as $\Delta \textrm{V}_\textrm{sys}$). FWHM denotes the Gaussian Full-Width at Half-Maximum.
    The other parameters are expressed in km/s. The value of $\Delta K_p$ is calculated by performing a Gaussian fit at the best fit $V_{\textrm{res}}$. The error bars are determined from the covariance matrix of the Gaussian fit.}
\end{table}

The $K_p-V_{\rm res}$ of other species of interested detected in other UHJs are presented in Appendix (see Fig\,\ref{Fig_no_detection}). These ones show positive correlation near the expected orbital position and may warrant follow-up observations. On averaged, most species have a blueshifted signal compared to the expected velocity (V$_{\rm res}$ = $\Delta$$V_{\rm sys} \approx -2.9$\,km.s$^{-1}$). The orbital velocity is also lower than expected {($\Delta$$K_{p} \approx -20.5\,$km.s$^{-1}$)}.
%----------------------------------------------------------------- 

\subsection{Trails of the signal in CCF maps} \label{result_trail}

\begin{figure}[!thbp]
\centering
  \includegraphics[width=8cm]{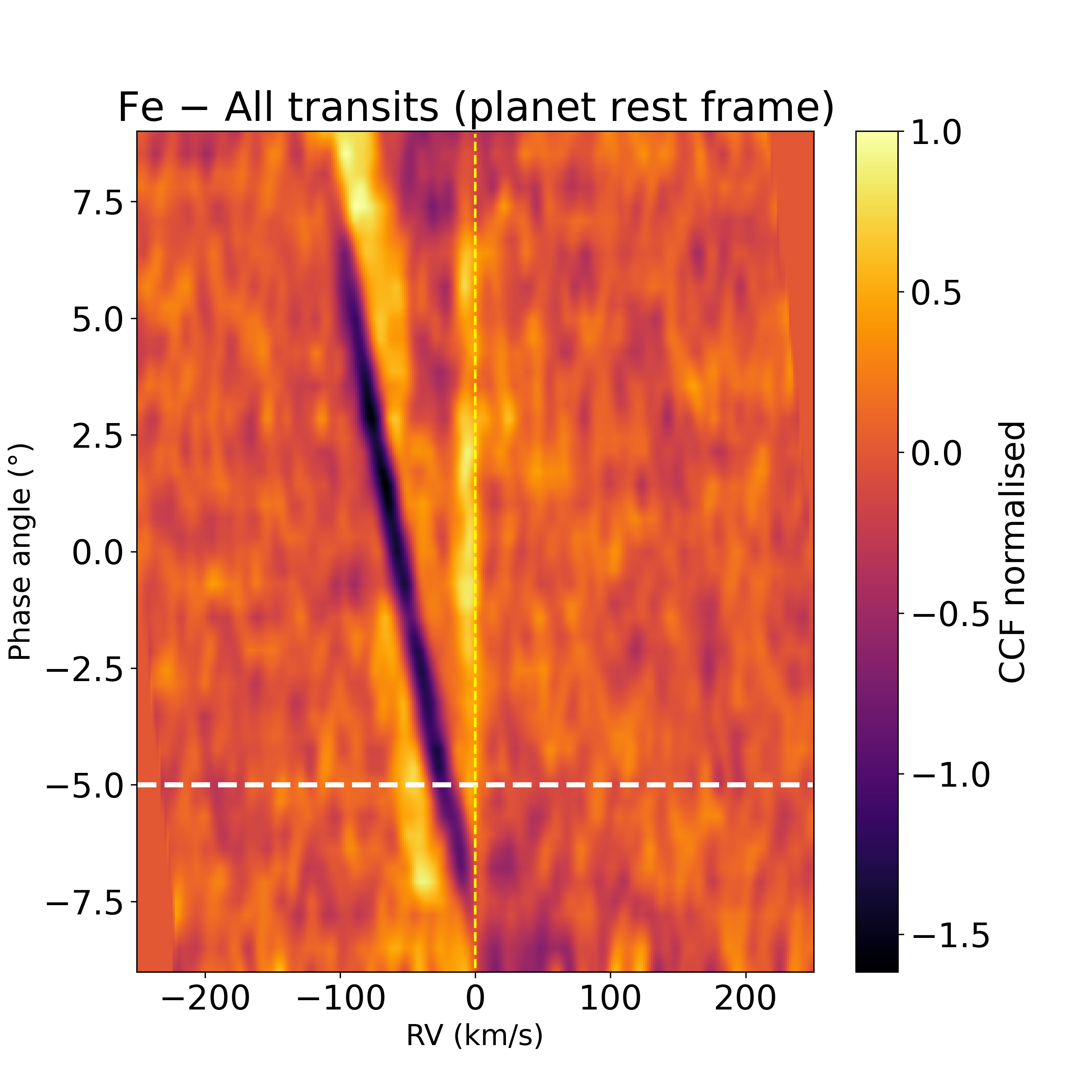}
  \caption{Cross-Correlation maps for iron in the planetary rest frame. The yellow dashed lines of the left panel represent the trace of the planetary signal at the expected $K_p-V_{\rm res}$. The yellow dashed line of the right panel represents the position of the planetary signal if the atmosphere is static. {The white dashed line represent the ingress part if the transit where we cannot distinguish the planetary signal (positive signal near 0 km/s) and the Doppler shadow (negative and positive signal in diagonal from -130 to 0 km/s).}
          }
     \label{Fig_doppler_shadow}
\end{figure}

Fig.\,\ref{Fig_doppler_shadow} presents { the time-variation of the CCF calculated with an iron template and shifted into the planetary rest frame}. The dashed yellow line on the right underlines the expected position of the planet if the atmosphere is static and homogeneous.
As the planet is misaligned, the Doppler shadow effect (already observed in \citealt{2021A&A...651A..33C}) only affects the planetary signal at the beginning of the transit and we have thus decided to not mask it nor to remove it. 

\begin{figure*}[!thbp]
\centering
  \includegraphics[width=18cm]{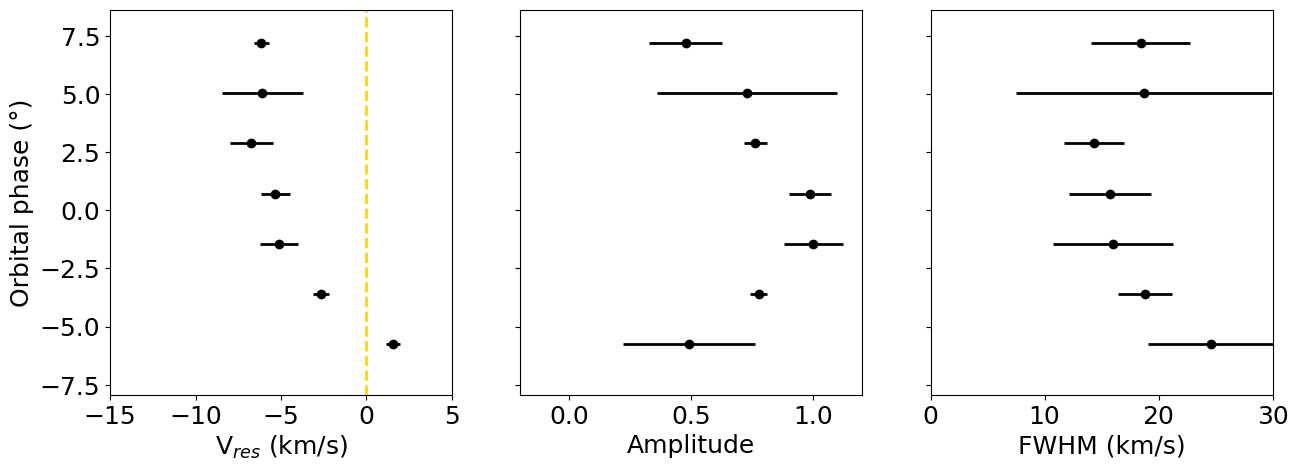}
  \caption{Position (left panel), Amplitude (central panel), and width (right panel) of the in-transit atmospheric CCFs Gaussian fit as a function of the orbital phase. The yellow dashed line of the left panel represents the expected position of the atmospheric track in the case of a static { and homogeneous} atmosphere.  
          }
     \label{Fig_trails_param}
\end{figure*}

The iron track shown in Fig.\,\ref{Fig_doppler_shadow} is slightly
shifted from the theoretical planetary velocity computed with the
orbital parameters of Table\,2. Previous high-resolution observations have demonstrated the ability to resolve time variations of this atmospheric track
\citep{2020Natur.580..597E,2021A&A...645A..24B}. We thus investigated this with our data. We binned the CCF similarly to \citet{2024PASP..136h4403W} to increase the planetary signal.
We divided the 25 in-transit frames observed in both datasets into nine bins. We use \texttt{scipy.optimize.curve\_fit} to fit a Gaussian to each bin between $\pm$ 10 km/s in the planetary rest frame. This gave better results than performing the fit across a broader range of velocities.
The results are shown in Fig.\,\ref{Fig_trails_param}. The velocity center of the atmospheric track changes with time (Fig.\,\ref{Fig_trails_param}, Left Panel), becoming more blue-shifted from around +1 km.s$^{-1}$ to around -8 km.s$^{-1}$. This is similar to what happens in the case of other UHJs like WASP-76b and WASP-121b \citep{2020Natur.580..597E,2021A&A...645A..24B}. 
The signal's amplitude reaches its peak during mid-transit, rather than at the beginning or the end.

\begin{figure*}[!thbp]
      %\includegraphics[trim = 3.5cm 0cm 4cm 1cm
      %,clip,width=\textwidth]{Figures/CCF_map/CCF_Trail.png}
      \includegraphics[width=\textwidth,trim = 3cm 0 2cm 0,clip] {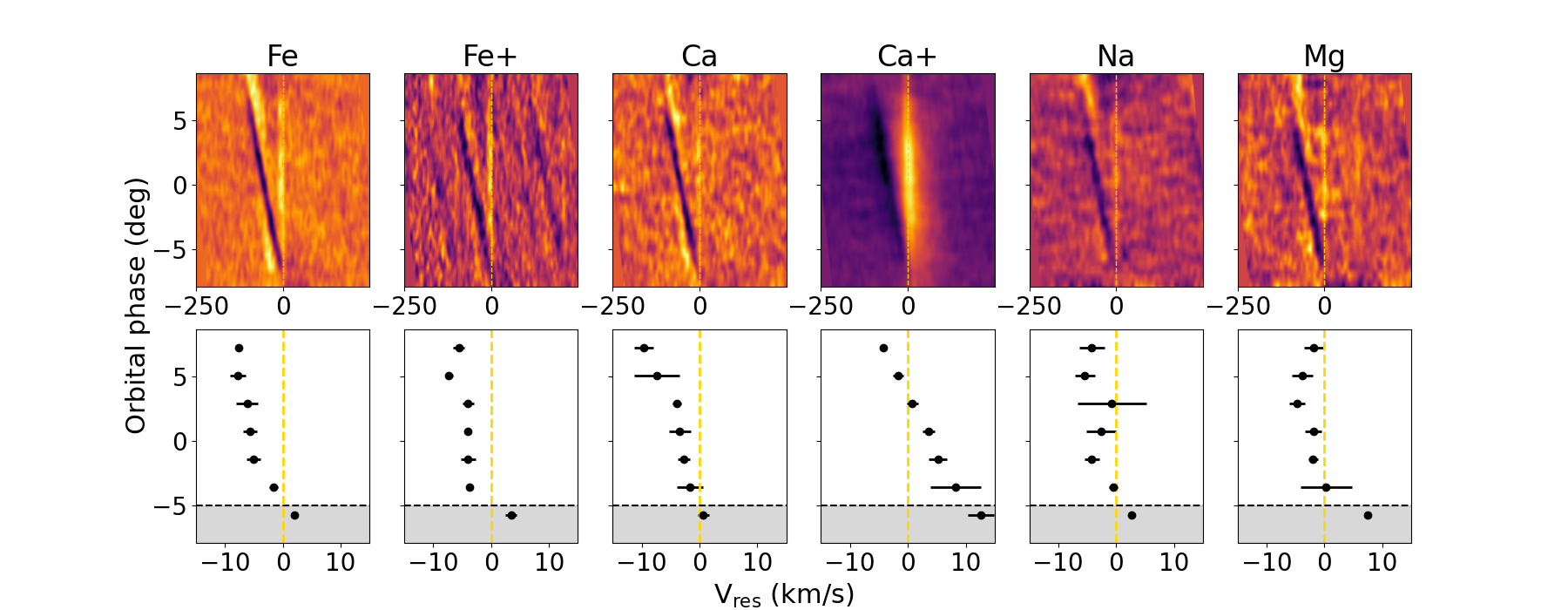}
      \caption{\textbf{Upper panel} : CCFs map of TOI-1518\,b for 6 different species. 
      \textbf{Lower panel}: Position of the maximum Gaussian fit of the CCF maps of the upper panel. The yellow dashed line of the left panel represents the expected position of the atmospheric track in the case of a static { and homogeneous} atmosphere.
              }
         \label{Diff_Trail}
\end{figure*}

We further detect the time-varying trace of the CCF for six different species: Fe, Fe+, Ca, Ca+, Na, Mg (Fig. \ref{Diff_Trail}). All species show a qualitatively similar behavior to the Fe trace, with a blueshift over time. 

\begin{figure*}
    \centering
    \includegraphics[width=18cm]{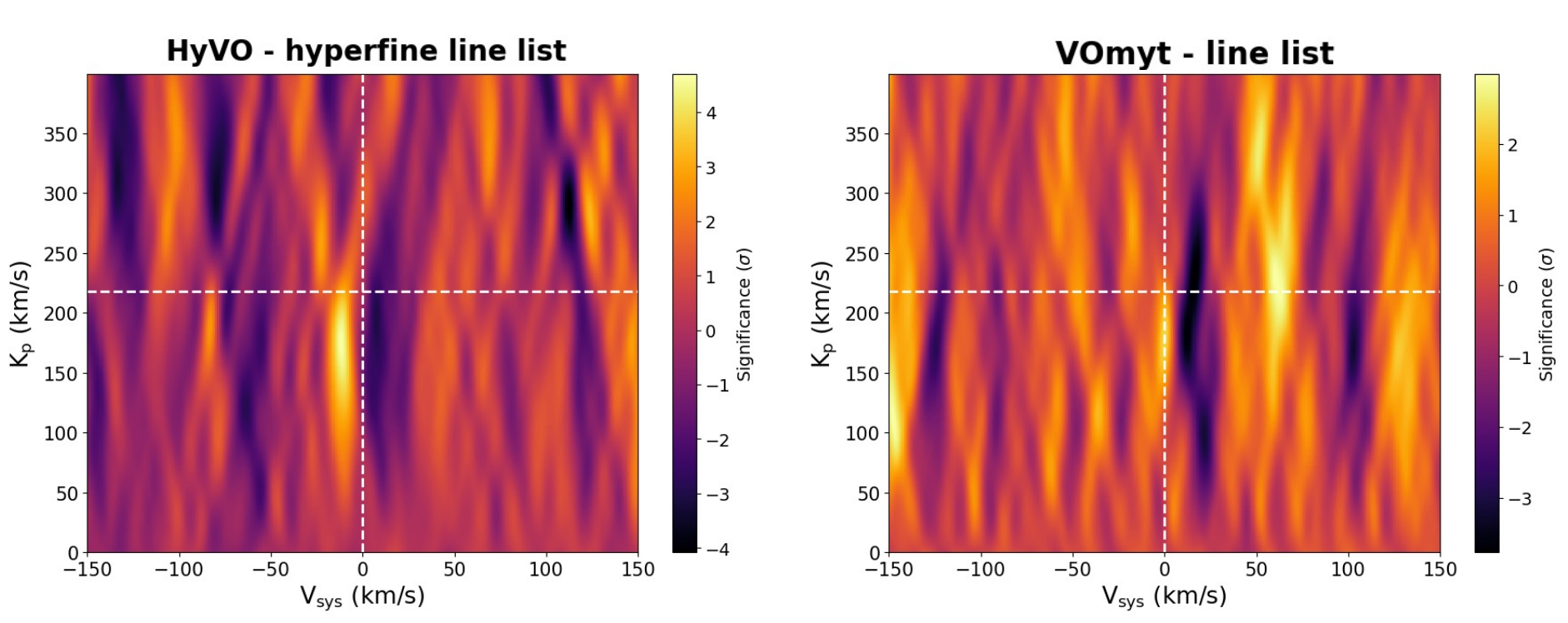}
    \caption{$K_p-V_{\rm res}$ map of VO in TOI-1518\,b with two different line lists. Left panel: CCF made with HyVO line list \citep{2024MNRAS.529.1321B}. Right panel: Same but with VOmyt line list \citep{2016MNRAS.463..771M}.}
    \label{fig:VO_comp2}
\end{figure*}

\subsection{Discussion} \label{disc_CCF}

The equilibrium temperature of TOI-1518\,b (2492\,K, \citealt{2021AJ....162..218C}) is lower than WASP-189\,b (2641\,K, \citealt{2018arXiv180904897A}) and higher than WASP-76\,b (2228\,K, \citealt{2020Natur.580..597E}), two recently UHJs observed with MAROON-X \citep{2023Natur.619..491P,2023A&A...678A.182P}. Then, each species observed in both planets is expected to be present in TOI-1518\,b. The detection of iron, manganese, chromium, vanadium, magnesium, calcium, and sodium is thus consistent with previous observations. The non-detection of potassium is due to the overlap of the telluric water lines with the Doppler-shifted potassium lines. This is the consequence of an unfortunate systemic velocity and barycentric velocity during these two observations. The detection of titanium in TOI-1518\,b, while it was not present on the cooler WASP-76\,b but present on the hotter WASP-189\,b, might be a sign that there is a trend of titanium abundance with temperature, possibly linked to the formation of TiO or to the nightside condensation of titanium. Whereas TiO would be expected in TOI-1518\,b, we are not able to find it.
 
The detection of VO was made with a significance of 4.9\,sigma using the newly released HyVO line list this year \citep{2024MNRAS.529.1321B}. VO is notoriously difficult to detect in exoplanet atmospheres; however, \citealt{2023Natur.619..491P} demonstrates the feasibility of such detections when employing a more accurate line list. For WASP-76b, the VOmyt line list was useful for the detection of VO. Fig.\,\ref{fig:VO_comp2} illustrates that this line list was unsuccessful in retrieving VO signals in the observations of TOI-1518\,b. This discrepancy can be attributed to a stronger VO signal in WASP-76b, along with the use of three transit observations compared to only two for TOI-1518\,b.
In this study, we demonstrate the superior capability of the HyVO line list, which successfully detected VO (see Fig.\,\ref{fig:VO_comp2}). The detected signal is blueshifted by as much as -12.11 $\pm$ 0.22 km/s (see Table\, \ref{tab_KP_Vsys}), which is significantly higher than the maximum blueshift observed for other species in TOI-1518\,b, recorded at -5.16 $\pm$ 0.57 km/s. Currently, it remains uncertain whether this difference is due to a shift in the line list itself or a physical shift in the atmosphere, potentially stemming from the varying localization of VO compared to metals and ions in the atmosphere of TOI-1518\,b. The presence of strong optical absorbers such as VO and TiO in the atmosphere of UHJs is crucial for thermal inversion phenomena \citep{2008ApJ...678.1419F}. This finding underscores the significance of utilizing more accurate line lists for the detection of these molecules, especially in cases where the signals may be weaker than those observed in WASP-76b. Further studies employing this new line list could reveal the presence of VO in targets where it was previously undetectable, as highlighted in \cite{2021A&A...645A..24B}.

Detecting Fe/Fe+, Ca/Ca+, Ti/Ti+, and potentially V/VO raises questions. For example, Fe+ is expected more in the hotter dayside atmosphere, whereas Fe should be more present on the cooler limbs and nightside. However, the tracks of Fe and Fe+ seen in Fig.~\ref{Diff_Trail} show a similar blueshifting trend, meaning that they are likely probing similar atmospheric regions. 

Additionally, the iron trail of TOI-1518\,b observed in Fig.\,\ref{Fig_doppler_shadow} and Fig.\,\ref{Fig_trails_param} follows a comparable trend than the ones previously observed in WASP-76\,b \citep{2020Natur.580..597E}, WASP-121\,b \citep{2021A&A...645A..24B} and WASP-189\,b \citep{2023A&A...678A.182P} with a signal becoming more blueshifted with the transit. 

One of the main uncertainties of this work is the impact of the PCA step on the planetary signal. Fig.\,\ref{Diff_comp} shows the iron K$_{\rm{p}}$-V$_{\rm{res}}$ map for three components removed on the left, and on the right, it shows the maximum of the K$_{\rm{p}}$-V$_{\rm{res}}$ map in the function of the number of PCA components removed for three different boxes where measured the standard deviation of the map.  The effect of PCA on high orbital velocity residuals is less than that on low velocity residuals, so calculating the standard deviation in the red box will underestimate the S/N. Conversely, the effect of PCA on low velocity residuals is much stronger, so the residuals will be smaller than those for high velocity, and the S/N will be overestimated. We then decided to calculate the standard deviation from the blue box, which mitigates this effect by taking the residuals at each velocity, but still far from the RM residuals or planetary signals, to avoid misinterpreting the standard deviation of the residual. The same method was used for each species observed.

Several K$_{\rm{p}}$-V$_{\rm{res}}$ maps, such as Fe, Mg, Ca, or Cr, show a parasitic signal at K$_{\rm{p}}\approx 90$ km/s and V$_{\rm{res}}\approx -60$ km/s. This signal is a residual of the Rossiter-McLaughlin effect observed in the CCF map in Fig. \ref{Fig_doppler_shadow}. The Rossiter-McLaughlin effect presents an anti-correlation signal at negative $K_p$. Therefore, it is not shown here, but the residual positive correlation visible in yellow around the anti-correlation signal in the CCF map is the parasite signal observed in the K$_{\rm{p}}$-V$_{\rm{res}}$ diagram mentioned previously. Possible biases due to Rossiter Mc-Laughing effect are present at phase below -5 degree and are represented in grey on the trail map of Fig. \ref{Diff_Trail}.
The positive correlation at phases above 5 degree is far from the planetary signal 5 ($< -50$ km/s).
For the case of Mg, a signal is also visible at expected K$_{\rm{p}}$ but V$_{\rm{res}}$ = 100 km.s$^{-1}$. This is due to a strong Fe+ line in the Mg triplet. This is also visible in the CCF maps of both species, where the residual signal of the other can be observed as highlighted in Fig.\,\ref{Fig_Fe_Mg}. 

Fig.\,\ref{Fig_Gauss_fit} highlights some limitations of the Gaussian profiles used to parameterize the iron trail of TOI-1518\,b as most one-dimensional CCF do not follow a simple Gaussian profile. Then we decided to center the fitted Gaussian profile on the maximum of the 1D-CCF even if this resulted in a misestimation of the FWHM and probably of the error bars of the measured  V$_{\textrm{res}}$. Another uncertainty might be due to the PCA applied to the data. Fig.\,\ref{Fig_nPCA} presents the iron trail of TOI-1518\,b with different numbers of principal components removed. The square root of the variance of the Doppler shifts across all numbers of removed
components, $\sigma$PCA has been added to the uncertainty quoted in the covariance matrix of the Gaussian fit obtained from \texttt{scipy.optimize.curve\_fit}. 
The PCA does not change the transit trend even when many components are removed for all species except Ca+ (Fig.\ref{Fig_comp_trails}). The PCA is essential for detecting faint species. To maintain consistency, we removed the same number of components (three) from all species, even for those where it might not have been necessary. The only exception is Ca+, as the PCA significantly impacts the signal, even when fewer components are removed, due to its very high signal strength. Therefore, we decided not to apply PCA to Ca+ in order to achieve a more robust analysis of its trail. 

\section{Comparison with global circulation models } \label{sec:GCM}

\begin{figure*}[!thbp]
   \centering
   \includegraphics[width=16cm]{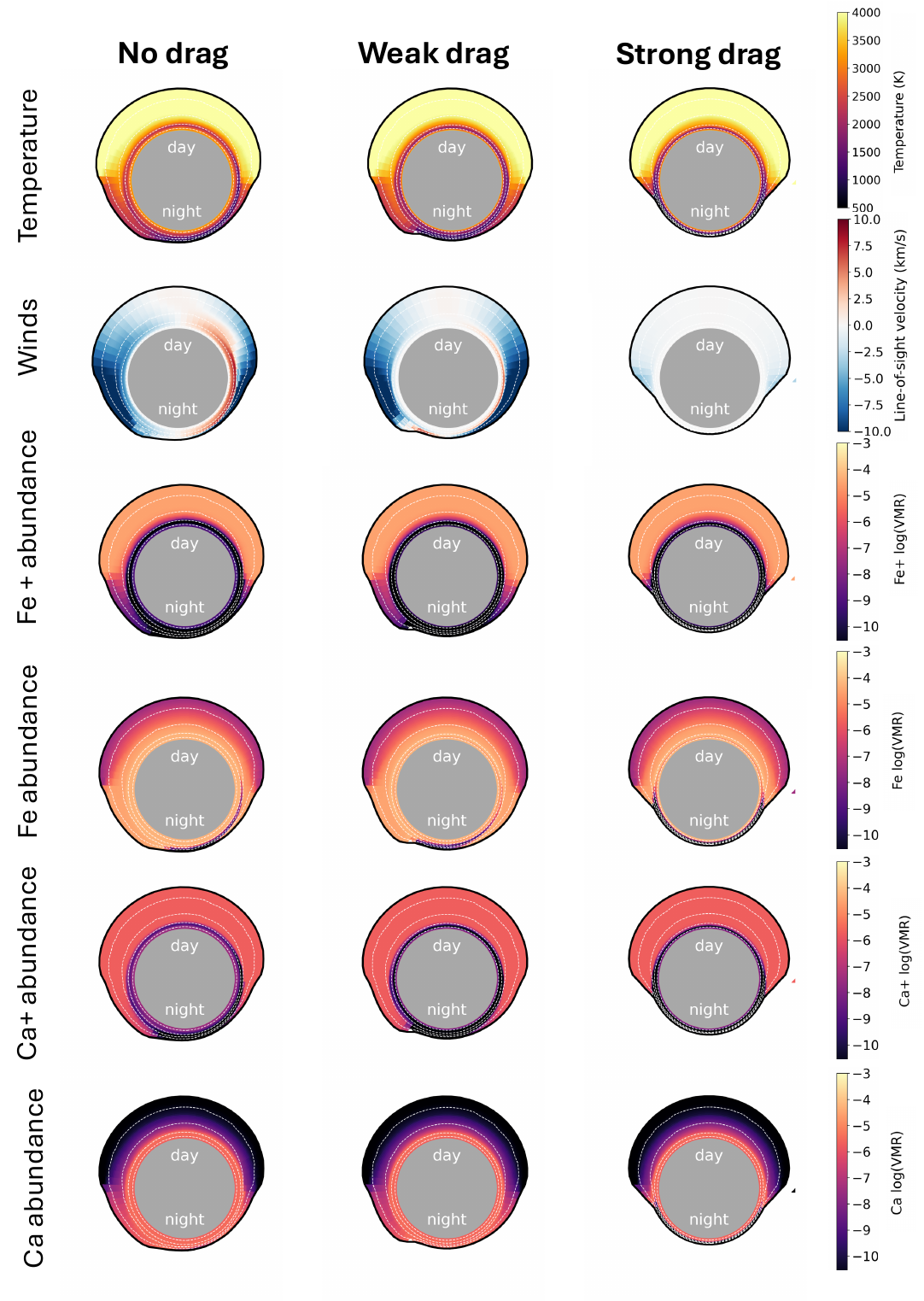}
   
      \caption{Overview of three out of the five GCM models of TOI-1518\,b considered in this work. The model in the first column is the drag-free model, while second and last column consider respectively weak ($\tau_{drag} = 10^6$s) and strong ($\tau_{drag} = 10^3$s) drag effects. Each panel shows the planet's equatorial plane, with the relative size of the atmosphere inflated for visualization purposes. From top to the bottom, the rows show the temperature structure, the line-of-sight velocities due to winds (at mid-transit),the spatial distribution of Fe+, Fe, Ca+ and Ca respectively. The white dashed contours in each plot represent isobars with pressures P = {$10^{1}, 10^{-1}, 10^{-3}, 10^{-5}$} bar}
         \label{fig: GCMs}
\end{figure*}

To understand the physics behind the six time-resolved absorption trails we observed, we now compare our data with a suite of GCMs.

\subsection{Model description} \label{GCM_explanation}

We consider five SPARC/MITgcm models of TOI\,1518\,b, three are presented in Fig.\,\ref{fig: GCMs}. The SPARC/MITgcm was initially introduced by \citet{2009ApJ...699..564S}.
It has been widely used to study the atmospheric physics and chemistry of (ultra-)hot Jupiters \citep{2010ApJ...709.1396F,2013ApJ...762...24S,2013ApJ...767...76K,2018A&A...617A.110P}. Our cloud-free models of TOI-1518\,b are based on work by \citet{2024MNRAS.528.1016T}. { As in previous works \citep{2013ApJ...762...24S,2016ApJ...821...16K,2018haex.bookE.116P}, we parameterize several source of dissipation (including turbulent mixing\citep{2010ApJ...725.1146L}, shocks \citep{2012ApJ...761L...1H} , hydrodynamic instabilities \citep{2016A&A...591A.144F}, or magnetic drag\citep{2010ApJ...719.1421P,2022AJ....163...35B}) through a Newtonian relaxation term in velocity applied to the momentum equation. The relaxation timescale, called the drag timescale hereafter, ranges from $\tau_{drag} = 10^3 $s (strong drag) $\tau_{drag} = 10^6$s (weak drag).}

These GCMs account for heat transport due to H2 dissociation and recombination (e.g., \citet{2018ApJ...857L..20B}; \citet{2018RNAAS...2...36K};\citet{2019ApJ...886...26T}; \citet{2021MNRAS.505.4515R}). H2 thermally dissociates on the dayside, after which atomic hydrogen gets advected to the nightside, where it recombines into H2 and releases latent heat. When the atmospheric circulation is predominantly eastward, most of this heat is pumped on the evening limb, resulting in a temperature asymmetry between the eastern and western regions of the atmosphere (first and second column in Fig.\ref{fig: GCMs}).
Drag restores energy to the atmosphere. Increasing drag strength (i.e lowering  $\tau_{drag}$) slows down winds in the atmosphere hindering this model's heat transport. In strong drag cases, the temperature structure becomes symmetric, resulting in similar chemical compositions in the morning and evening limbs. (last column in Fig.\,\ref{fig: GCMs}). In the strong-drag model, there is only a day-to-night flow as the equatorial jet gets suppressed \citep{2013ApJ...762...24S}. Table\,4 summarizes some other important parameters of the two SPARC/MITgcm models. We refer to Table 1 in \citet{2024MNRAS.528.1016T} for the full list of opacities considered in their radiative transfer (which include species such as TiO and Fe driving thermal inversion on the planet dayside). All models were run at a horizontal resolution of C32, corresponding to roughly 128 cells in longitude and 64 in latitude. Before computing phase-dependent spectra of the GCMs with gCMCRT, we bin the outputs down to 32 latitudes and 64 longitudes, as in \citet{2021MNRAS.506.1258W,2023MNRAS.525.4942W,2024PASP..136h4403W}.

\begin{table}[!htbp]
    \centering
    \caption[]{Overview of some of the parameters of the GCMs described in Sect. \ref{GCM_explanation} (see Fig.\ref{fig: GCMs} for plots of the equatorial plane of each model and Fig.\,\ref{fig: GCMs_injected} for the limb planes).}
    \label{GCMs_input}
    \begin{tabular}{c c}
        \toprule
        \midrule
        \noalign{\smallskip}
       Parameter &  Value   \\
        \midrule
        \noalign{\smallskip}
      Orbital Period  & 1.6442 $\times 10^5$s (1.903 days) \\
        \noalign{\smallskip}
       Pressure range  & 200 $-$ 2$\times 10^{-6}$ bars \\
        \noalign{\smallskip}
      Radius at bottom & 1.3405 $\times$ 10$^8$m (1.875 R$_\textrm{Jup}$)  \\  
        \noalign{\smallskip}
     Gravity &  10.56 m/s$^2$  \\
        \noalign{\smallskip}
      Horizontal resolution & C32 \\
        \noalign{\smallskip}
      Vertical resolution & 53 layers\\
        \noalign{\smallskip}
      Metallicity and C/O  &  1 $\times$ solar \\
        \noalign{\smallskip}
      H/H$_2$ heat transport  &  $\checkmark$\\
        \noalign{\smallskip}
     Drag timescale  &   \{$\infty$,$10^3$s,$10^4$s,$10^5$s,$10^6$s\}\\
        \noalign{\smallskip}
      Radiative transfer  &  non-grey (see \citealt{2013ApJ...767...76K})\\
        \noalign{\smallskip}
        \bottomrule
    \end{tabular} \\ 
\end{table}

\subsection{Injection and cross-correlation maps}

\begin{figure*}[!thbp]
   \centering
       \includegraphics[width=19cm]{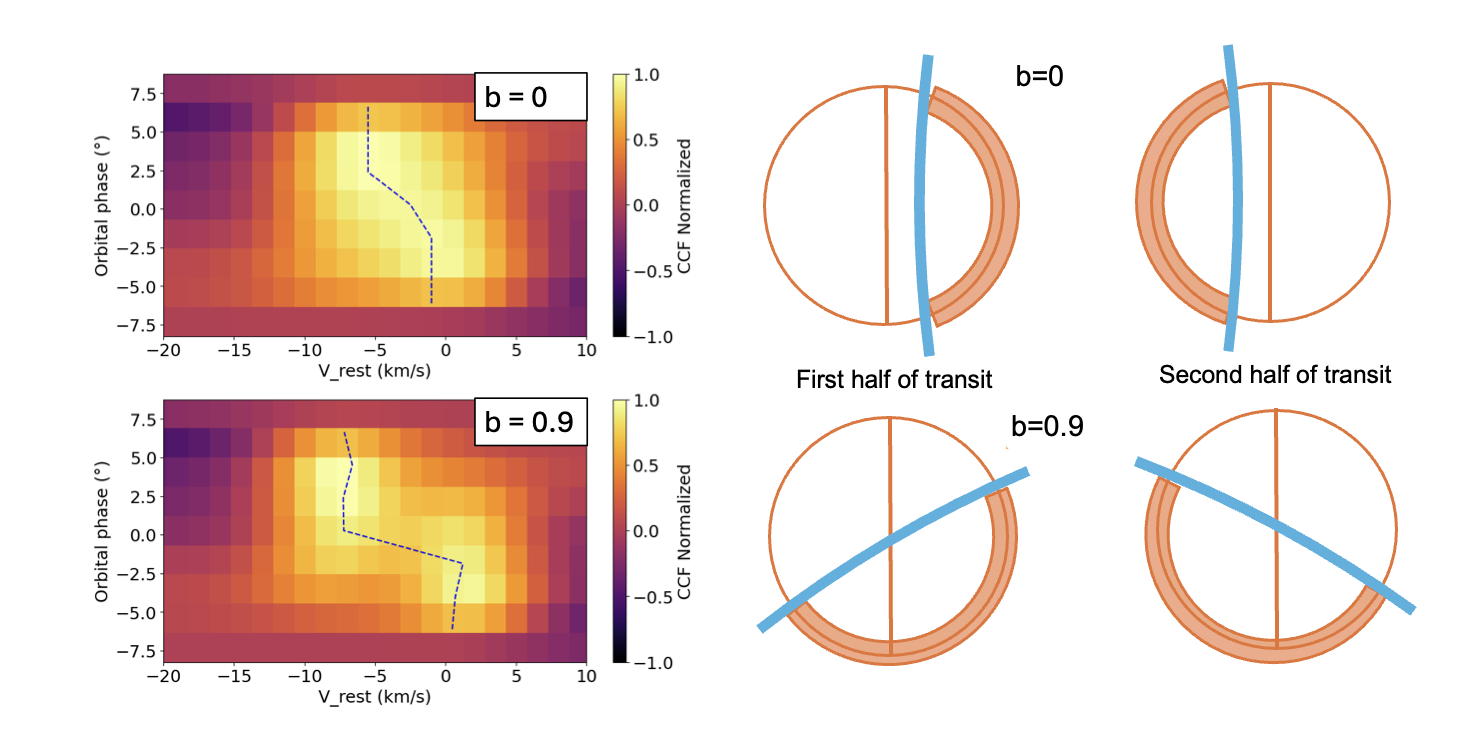}
      \caption{Importance of the impact parameter b with examples for b =0 on the top raw and b=0.9 on the lower raw. On the left, the CCF maps of the injected models within blue dashed lines, the Gaussian fit applied similarly to the data (Fig.\ref{Fig_trails_param}). On the right, the illuminated part of the limb during the first and second half of the transit is represented in a vertical slice, shown to scale.
              }
         \label{fig: GCMs_injected}
\end{figure*}

We compute phase-dependent transmission spectra of the five GCM models across the MAROON-X spectral range (between 490 and 920\,nm) using gCMCRT \citep{2022ApJ...929..180L}. The calculations account for Doppler shifts due to planet rotation and winds. Section\,2 of \cite{2023MNRAS.525.4942W} shows the radiative transfer and post-processing details. Therefore, we will only briefly summarize the gCMCRT setup for TOI-1518\,b analysis.
Before feeding the GCM outputs into gCMCRT, we map the atmospheric structures onto a 3D grid with altitude (instead of pressure) as a vertical coordinate. We account for the fact that each atmospheric column has a different scale height set by local gravity, temperature, and mean molecular weight. For each of the five TOI-1518\,b models, we simulate 25 spectra (equidistant in orbital phase) between phase angles $\pm$ 8 degrees, covering the in-transit part of our observations.  
At each orbital phase angle, gCMCRT simulates a transmission spectrum by randomly shooting photon packets at the part of the planet limb that is blocking the star and evaluating the optical depth encountered by each photon packet. Due to the geometry of TOI-1518\,b (Fig.\,\ref{FigGeometry}), the part of the planet that is illuminated differs from the case of WASP-121\,b where the impact parameter is near 0 (Fig.\,\ref{fig: GCMs_injected}, Right Panel). In this calculation, the code accounts for Doppler shifts imparted on the opacities by the radial component of the local wind vector and planet rotation \citep{2021MNRAS.506.1258W}.
The transit depth at a specific wavelength is calculated by averaging it over all photon packets. To accurately represent the shapes, depths, and shifts of the spectral lines, we use $10^5$ photon packets per wavelength. Since we do not explicitly account for scattering, the direction of propagation for the photon packets remains constant throughout the calculation. { The spectra are simulated at a native resolution of R = 300\,000, and then convolved down to the instrument resolution of $R_{\lambda} = 85\,000$, which differs from \cite{2023MNRAS.525.4942W, 2024PASP..136h4403W}.} We include the same set of continuum opacities and line species for the radiative transfer as in \cite{2023MNRAS.525.4942W}.
To properly compare the GCM spectra with the data, we need to inject the GCMs into the data and perform PCA, similar to what we did for the data. The different steps were as follows: 

\begin{itemize}
    \item Inject the spectra at each orbital phase into the three components that were removed during the data reduction process, using the $K_{\rm{p}}$ velocity value from Eq. \ref{eu_eqn}.
    \item Perform PCA on the data+GCM removing three components, the same number of components as for the data in the case of iron. 
    \item Subtract the post-PCA data from the combined data+GCM to isolate the GCM signal post-PCA and remove the noise.
    \item Do the cross-correlation with the same Fe template and methods used for Sect.\,\ref{sectionCCF}. 
\end{itemize}

The resulting CCFs for the case of the drag model are shown in Fig.\,\ref{fig: GCMs_injected} . The difference between the two models at different impact parameters b is also represented with the case of b = 0 on the upper row and b=0.9 on the lower row. 

We then interpolated the cross-correlation results over the same phase grid as for the observations of the two transits. We then performed the same analysis using the fit with a Gaussian profile, which allowed us to extract the position, amplitude, and FWHM of the model's signal.

\begin{figure*}[!thbp]
   \centering
      \includegraphics[width=\textwidth]{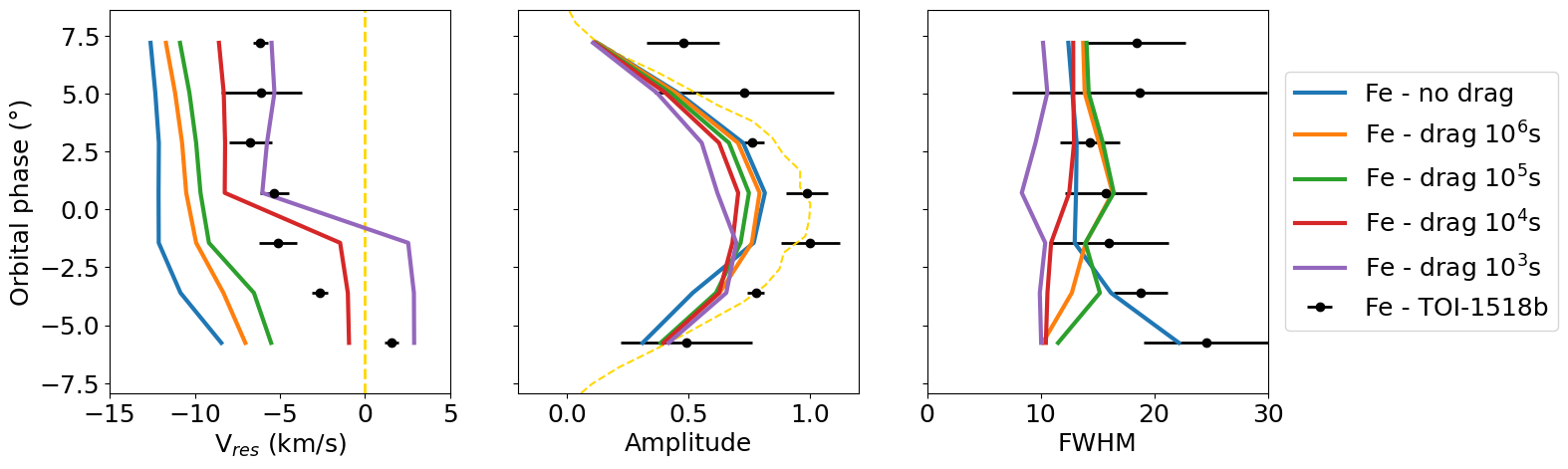}
      \caption{Same as Fig.\ref{Fig_trails_param} but with the results of the Gaussian fit for the different models from no drag effect to strong drag effect. In the central panel, the illuminated fraction of the planetary limb as a function of the orbital phase is represented in a yellow dashed line. 
              }
         \label{Fig:Gauss_model}
\end{figure*}

\begin{figure*}[!thbp]
   \centering
      \includegraphics[width=\textwidth]{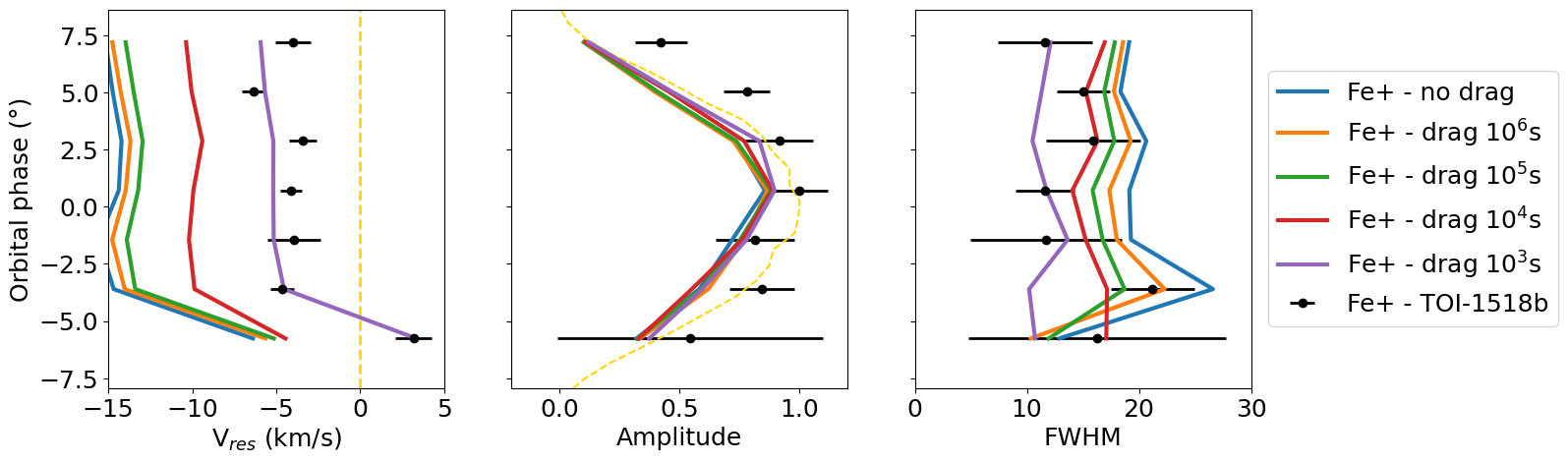}
      \caption{Same as Fig.\ref{Fig:Gauss_model} but with the results of the Gaussian fit for the Fe+ compared to models. 
              }
         \label{Fig:Fe+_model}
\end{figure*}

Figs.\,\ref{Fig:Gauss_model} and \ref{Fig:Fe+_model} show the CCF signals of Fe and Fe+ that we obtain for each of the models, with the real data plotted on top.

\begin{figure*}[!thbp]
   \centering
      \includegraphics[width=\textwidth,trim = 0cm 0 0cm 0,clip]{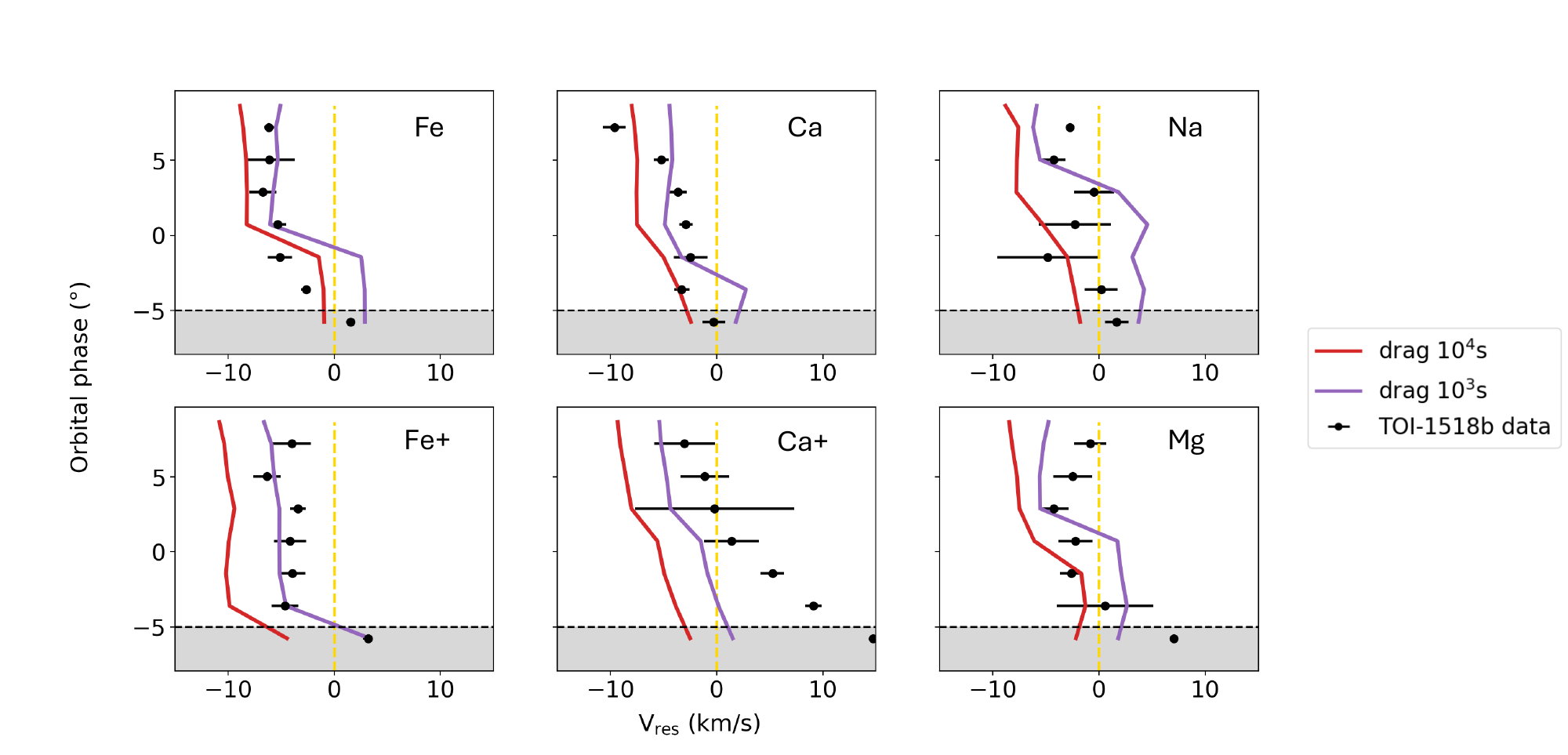}
      \caption{Same as lower panel of Fig.\ref{Diff_Trail} but with the results of the Gaussian fit for the stronger drag models included. 
              }
         \label{Fig:all_species_model}
\end{figure*}

\subsection{Discussion}

{ Fig.\ref{Fig:Gauss_model} (Left Panel) shows the Fe trails of the five GCMs simulations, along with the Doppler shifts of Fe measured in Sect. \ref{sectionCCF}. All models show a signal that is blueshifting with time. This behavior was previously observed by \citet{2020Natur.580..597E,2021A&A...645A..24B} and discussed in many papers \citep{2021MNRAS.506.1258W,2023MNRAS.525.4942W,2022ApJ...926...85S,2022AJ....163...35B,2023AJ....165..257B}. Because the planet rotates during the transit, the start of the transit probes the dayside of the leading limb and the nightside of the trailing limb. It is thus dominated by the hotter leading limb. The planet rotation red-shifts the leading limb by $\approx$ 3.5 km/s, which is compensated by the blue shift of the day-to-night winds. In the second part of the transit, the opposite happens and the signal is dominated by the trailing limb, which is blueshifted by rotation. The change from being dominated by the leading and the trailing limb during transit is responsible for the blueshifting of the signal with time. 

In the case of TOI-1518\, b's atmosphere the high impact parameter reinforces this blue shift. Indeed, as shown in Fig.\,\ref{fig: GCMs_injected}, because the planet is close to being grazing, we almost never probe the full atmospheric limb. The signal from the first half of the transit is therefore entirely due to the leading limb, with only a very small contribution to the trailing limb, and the opposite is true for the second part of the transit. As a consequence, we find that a given planet is expected to have a stronger blueshift with time when observed with a high impact parameter than if it has a low impact parameter.

Furthermore, the global shift of the planetary trace is linked to the day-to-night winds that blueshift both leading and trailing hemisphere signals. These day-to-night winds are strongly affected by drag. As shown in Fig.\ref{Fig:Gauss_model}, models without drag produce an iron signal that is too blueshifted compared to the observations. Overall, the planetary signal stands between the stronger drag models ($\tau_{drag} = 10^4$ and $10^3$s). When looking at the amplitude and at the FWHM of the signal, however, we see that the $\tau_{drag} = 10^4$s model is favored, because the $\tau_{drag} = 10^3$s provides a signal that is too small and not wide enough compared to data. 

If we now look at the Fe+ trace (Fig.\ref{Fig:Fe+_model}), we see that the $\tau_{drag} = 10^4$s overestimates the blue shift, whereas the $\tau_{drag} = 10^3$s ( in purple) is a much better match. For Fe+ the amplitude of the model is unaffected by the drag and the FHWM errorbars are too large to differentiate between the two strong drag models. The apparent contradiction between the Fe trace favoring the $\tau_{drag} = 10^4$s model and the Fe+ trace favoring the stronger, $\tau_{drag} = 10^3$s drag model could be an indication of the presence of ohmic drag in the atmosphere. Indeed, as shown by \citet{2022AJ....163...35B}, Ohmic drag strength strongly depends on planetary location, being stronger in the hotter dayside and at lower pressures. Because the Fe/Fe+ equilibrium is temperature dependent, the Fe+ signal naturally probes more into the hotter planetary dayside than the Fe signal, where the temperatures are hotter and magnetic drag effects are expected to be larger. 

Fig.\ref{Fig:all_species_model} shows the trails for the 6 species of Fig.\ref{Diff_Trail} with the signals of the stronger drag models. For Ca, Na, Mg, and Ca+ the signal straddles between both these GCMs, as in the case of Fe. We therefore confirm that the strong drag scenario is preferred by the signal of all the other species. A similar agreement with an intense drag model atmosphere was already observed in the case of WASP-121\,b by \citealt{2024PASP..136h4403W}, pointing out that strong drag is common in ultra-hot Jupiter atmospheres. \\ 

The trail of the Ca+ is significantly different from the ones of the other species, as it starts strongly red-shifted. This is clearly not captured by the model. We calculated the contribution functions of our model, and find that the CCF of most species (including Ca, Fe and Fe+) probe the $10^{-3}$ and $10^{-5}$ bar layers whereas the Ca+ probes layers up to $10^{-8}$. We fitted directly the resolved individual Ca+ lines using a method similar to that of \citet{2024A&A...685A..60P} and found a transit depth corresponding to an effective radius of $1.97 \pm 0.04\, \textrm{R}_p$. This value is similar to the planetary Roche radius (R$_{\textrm{Roches}} \simeq 1.96 \textrm{R}_\textrm{p}$ at Lagrange Point 1 following \citet{2003ApJ...588..509G}). As a consequence, we believe that the Ca+ lines likely probe the outflow of the planet. This region shows a different atmospheric flow than the deeper regions, with an important contribution from planet-star interactions, not modeled in our GCM. Our results for Ca+ are similar to the recent observations of H-alpha for WASP-121b \citep{2025Natur.639..902S}.}

{The CCF of the GCM, presented in Fig.\,\ref{fig: GCMs_injected}, exhibits a double-peak structure with one centered near 0 km/s and another around -7 km/s. This feature arise from the contribution of both atmospheric limb. Fitting this complex signal with a simple Gaussian profile is inherently limiting, as a single-peaked function cannot accurately capture the asymmetric nature of the CCF. Parameters such as the FWHM and amplitude may be misrepresented, potentially leading to an incomplete interpretation of the atmospheric dynamics.
Finally, for vanadium oxide, the predictions from the GCMs do not match the observed signal observed in Fig. \ref{fig:VO_comp2}. The GCMs estimate a change in K$_{\rm{p}}$ of about $\pm$ 5 km/s and a V$_{\rm{res}}$) ranging from -6 to -13 km/s (decreasing with decreasing drag) when the signal observed in the data is at $\Delta K_{\rm{p}}$ around -33 km/s and V$_{\rm{res}}$ around -13km/s. }

\section{Retrieval analysis}

After detecting the species shown in Sect.\,\ref{sectionCCF} thanks to cross-correlation analysis, the next step is to explore the abundances of these species in comparison to solar values (This comparison is possible due to the solar metallicity of the host star). One approach developed in \citealt{2019AJ....157..114B} is to use a Bayesian atmospheric retrieval framework with high-resolution cross-correlation spectroscopy (HRCCS) that relies on the cross-correlation between data and models for extracting the planetary spectral signal. This approach allows the characterization of many atmospheres of UHJs and puts constraints on abundances \citep{2021Natur.598..580L, 2021ApJ...921L..18K,2023AJ....165....7K,2023AJ....165...91B}.

\subsection{CHIMERA code}

Following the above method, we applied the \citealt{2019AJ....157..114B} cross-correlation-to-log-likelihood retrieval framework to derive the molecular volume mixing ratios and the temperature layer we are probing. For the retrieval process, we used the CHIMERA “free-retrieval” \citep{2013ApJ...775..137L, 2015ApJ...814...66K} paradigm, which assumes constant-with-altitude gas mixing ratios and uses a simple isothermal T-P profile, which is a good enough approximation to retrieve abundances with MAROON-X as shown by \citealt{2023Natur.619..491P}. We use a similar setup as \citealt{2021Natur.598..580L} but with a wavelength range and a choice of chemical species adapted to the MAROON-X bandpass. Additionally, we use a native resolution for the radiative transfer of 500.000, that is downgraded to the MAROON-X 85.000 resolution after instrumental and rotational broadening and orbital Doppler shift. The retrieval parameters specific to our analysis and their prior ranges are provided in Table \ref{tab:Prior_chimera}. A more detailed description of the high-resolution GPU-based radiative transfer method and log-likelihood implementation within pymultinest \citep{2009MNRAS.398.1601F, 2014A&A...564A.125B} is given in \citealt{2021Natur.598..580L}. 

\begin{table}[!th]
    \centering
        \caption{Parameters and corresponding priors used in the retrieval analysis with CHIMERA for the study of TOI-1518\,b.}
    \begin{tabular}{c c c}
        %\hline
        \toprule
        \noalign{\smallskip}
        Parameter & Description & Prior \\
        \noalign{\smallskip}
        %\hline
        \midrule
        \noalign{\smallskip}
        T0 & Isothermal temperature & 1500 - 4500 K \\
        \noalign{\smallskip}
        xRp & Scaled radius of the planet & 0.5 - 1.5 \\
        \noalign{\smallskip}
        Mp & Mass of the planet & 0.5 - 2.3 \\
        \noalign{\smallskip}
        $log_{10}($Pc) & Continuum & -6 - 0\\
        \noalign{\smallskip}
        Kp & Planet Orbital velocity & 100 - 300 km/s \\
        \noalign{\smallskip}
        Vsys & Systemic Velocity & -100 - 100 km/s \\
        \noalign{\smallskip}
        $log_{10}($a) & Model scaling factor & -2 - 2 \\
        \noalign{\smallskip}
        $log_{10}($H-) & Continuum & -12 - 0 \\
        \noalign{\smallskip}
        $log_{10}(\chi_i$) & log gas volume mixing ratio & -12 - -1.5 \\ 
        \noalign{\smallskip}
        \bottomrule
        \noalign{\smallskip}
    \end{tabular}

    \label{tab:Prior_chimera}
\end{table}

\subsection{Results of the retrieval analysis} \label{sectionRetrieval}

We combined the blue and red arm MAROON-X data for our retrieval for both transits. We included most of the species detected in Sect.\,\ref{sectionCCF} $+$ TiO and abundance proxies for the H bound-free and free-free continua. However, some species detected in the Cross-Correlation analysis were not included in the retrieval framework due to a lack of opacities files in the right format, like Ba+, Si and Mn.

{The mass prior is set using the SOPHIE observation results (see Table \ref{table_param_fit}) as a free parameter that follows a Gaussian distribution with Mp = 1.83 M$_\textrm{Jup} \pm 0.47$.} For this analysis, we consider the two detectors independently as for the CCF, and we add the likelihood from the red and the blue detector analysis and then from both transit. For the retrieval analysis, we used three number of principal components as for the CCF analysis. As shown before, the PCA does not strongly affect the planetary lines, apart from the ones of the Ca+. However, because the retrieval considers all species at the same time, we could not have different PCA numbers per species. {The results of the measured abundances are summarized in Fig.\,\ref{multi_species_retriev} and the whole corner plot is given in the Appendix.(Fig.\,\ref{fig:Full_retrieval})}. 

\begin{figure*}[!thbp]
   \centering
      \includegraphics[width=18.5cm]{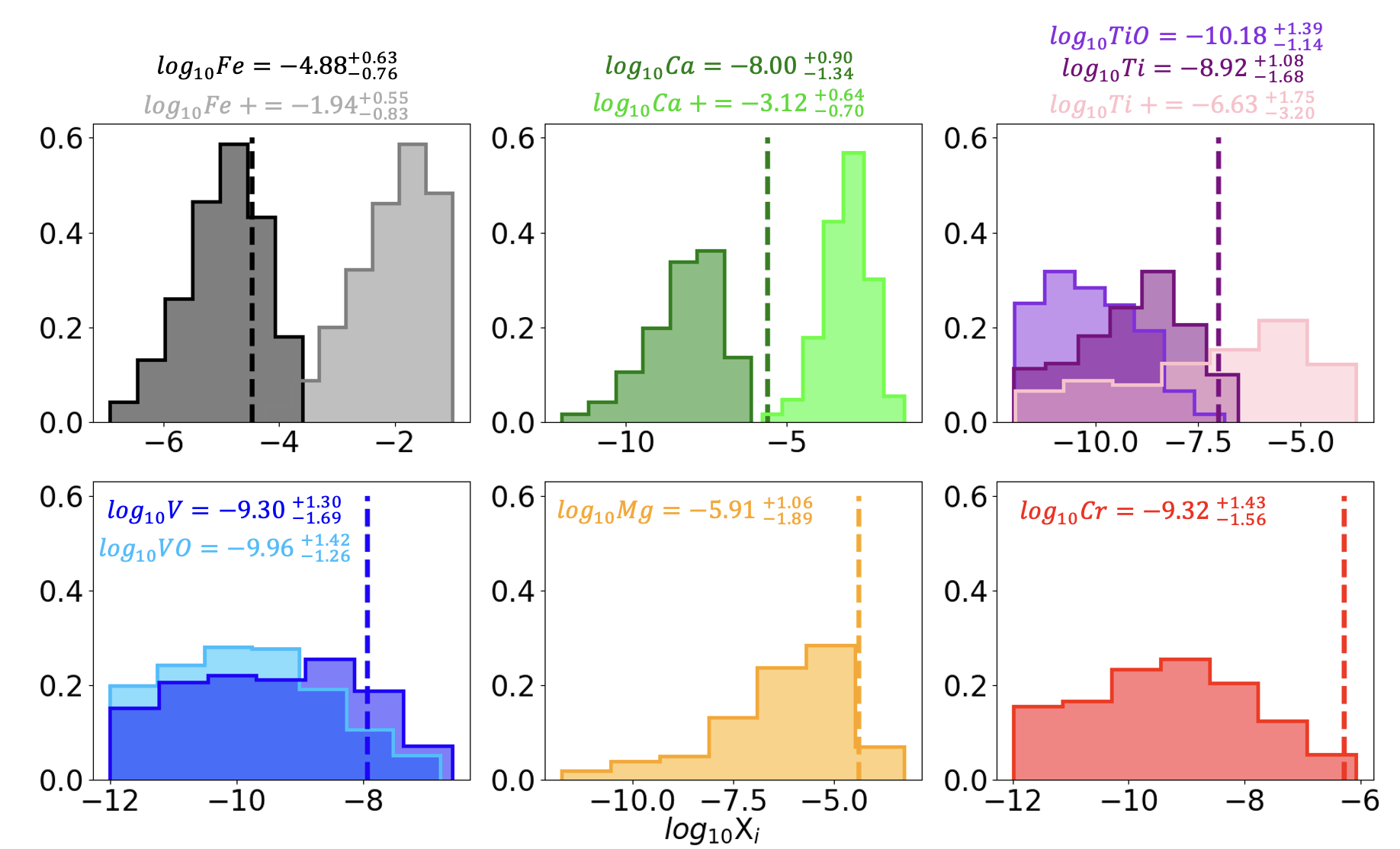}
      \caption{Histograms of the measured abundances of different elements in TOI-1518\,b’s atmosphere. Proto-solar value of each element is represented in  dashed lines (from \citealt{2019arXiv191200844L}).} %All error bars represent 1\,$\sigma$ uncertainties.} %Right panel : Histograms of the measured refractory abundance ratios in TOI-1518\,b’s atmosphere relative to proto-solar
      \label{multi_species_retriev}
\end{figure*}

\subsection{Discussion of the retrieval analysis}

The K$_{\rm{p}}$ and V$_{\rm{res}}$ distribution of the retrieval analysis(Fig.\,\ref{fig:Full_retrieval}) are consistent with the iron signal detected in Sect. \ref{sectionCCF}. {The measured V$_{\rm{res}}$ ($-3.05^{+0.62}_{-0.70}$ km/s) and K$_{\rm{p}}$ ($181.99^{+8.18}_{-8.16}$ km/s)} also agree with the detection of \cite{2021AJ....162..218C} ($\Delta\textrm{V}_{\textrm{sys}} = -2.06 ^{+2.00}_{-4.00}$ km/s and K$_{\rm{p}}$= $157^{+44}_{-68}$ km/s).  {With a $\Delta \rm{K_p}$ of approximately -25 km/s compared to the $\rm{K_p}$ derived from the SOPHIE data, and a V$_{\rm{res}}$ around -3 km/s, the retrieval is also capturing the blueshift of the signal as a function of the planet's orbital phase, as detailed in Sect. \ref{sec:GCM}}. We observed that the velocity parameters obtained are influenced predominantly by the strongest absorber, particularly Fe. 

{ Our model retrieves an iron abundance of $ {\log_{10}}$Fe=$-4.88^{+0.63}_{-0.76}$. This corresponds to 0.07 to 1.62 solar enrichment in iron. This remains consistent with a solar (and then stellar) metallicity. We have checked that the iron abundance was robust to different assumptions in the retrieval, and that it was not changing when different other species were added.}

{ The retrieval results for the other chemical species are more surprising and highlight the  difficulty of atmospheric retrievals for transit spectroscopy at high-spectral resolution. First, the ionized species, such as Fe+ and Ca+ have retrieved abundances that are much higher than solar. These high abundances, up to $ {\log_{10}}$Fe=$-1.94^{+0.55}_{-0.83}$ are unlikely in a gas giant atmosphere. Instead, we believe that these could stem from a lack of flexibility in the forward model. Indeed, ionized species are likely to probe different parts of the atmosphere (both vertically and longitudinally) than neutral species, because the spatial repartition of neutral and ionized forms of the same species is anti-correlated. This can lead to signals at different temperatures and K$_{\rm{p}}$ and V$_{\rm{res}}$ (e.g. Table~\ref{tab_KP_Vsys}). However, given that the temperature, K$_{\rm{p}}$ and V$_{\rm{res}}$ for all species is determined by the neutral iron lines, which represent the strongest signal, the abundances of the other species are likely biased to compensate for the K$_{\rm{p}}$/V$_{\rm{res}}$ offset.}

{For lines with a strong Kp/Vsys shift and small opacities, such as Cr, V, Ti or VO, the retrieval leads only to upper limits, which are all consistent with a solar enrichment.}

{For the neutral calcium and magnesium lines, the retrieval provides a detection with an abundance of $ {\log_{10}}$Ca=$-8^{+0.9}_{-1.34}$ and $ {\log_{10}}$Mg=$-5.91^{+1.06}_{-1.89}$, corresponding to 0.0002 to 0.034 times the solar abundance for Ca and 0.0004 to 0.339 times solar for Mg. The retrieved value for neutral calcium is significantly lower than that of iron, and it is, for now, unclear whether this is due to a physical effect (e.g. most Ca being into Ca+) or a bias in the retrieval. 
Finally, our results for VO deserve a special mention. }

\begin{figure*}[!thbp]
   \centering
      \includegraphics[width=18cm]{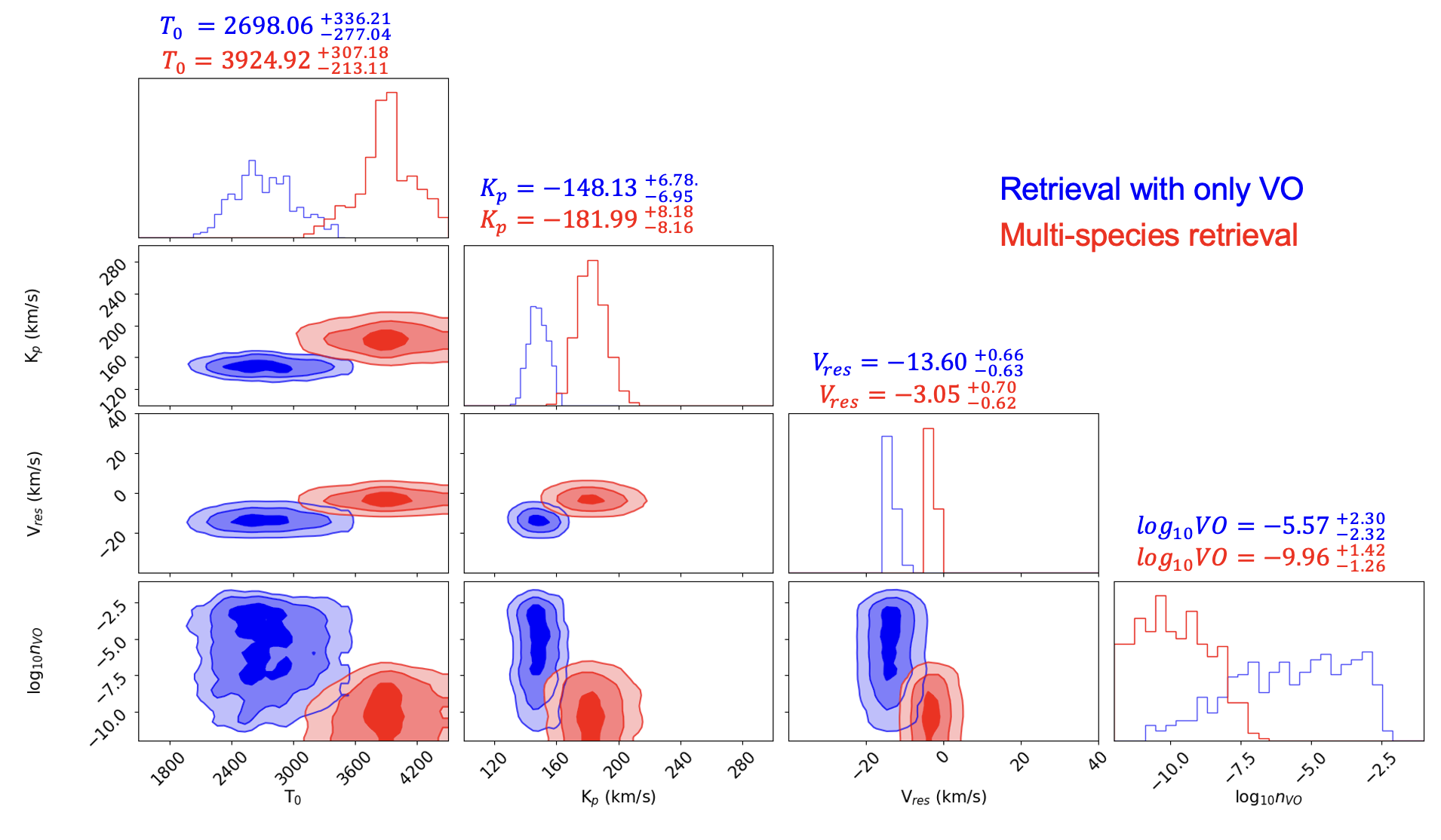}
      \caption{Likelihood distributions for temperature, $K_p$, V$_{res}$ and $log_{10}$VO derived from the multi-species retrieval (in red) and from retrieval with only VO inside (in blue). 
              }
         \label{VO_retriev}
\end{figure*}

{Despite using the latest VO line list, the multi-species retrieval did not successfully detect VO. To explore this issue further, we conducted a single-species retrieval focused solely on VO (see Fig. \ref{fig:Retriev_Vo_only}). This approach resulted in a detection with a temperature around 2700 K, compared to 3500 K in the multi-species retrieval, along with a different $K_{\rm{p}}-V_{\rm{res}}$ distribution (Fig. \ref{VO_retriev}). The measured value of VO in such a framework is $ {\log_{10}}$VO$=-5.57^{+2.3}_{-2.32}$, corresponding to an enrichment of 1.10 to 45708.81 times solar.  The discrepancy in the $K_{\rm{p}}-V_{\rm{res}}$ distribution resembles the patterns observed in the $K_{\rm{p}}-V_{\rm{res}}$ map (see Table \ref{All_Kp_Vrest}). This difference could be attributed to either dynamical factors—if VO probes specific regions of the atmosphere—or inaccuracies in the positioning of the lines within the VO line list. This situation underscores the limitations of the current framework in accurately deriving abundance for species that are settled in different $K_{\rm{p}}-V_{\rm{res}}$ distribution.

Further observations with bluer or redder instruments to detect molecules or ionized species are needed to determine why such elements are subsolar.
Moreover, using a non-isothermal profile instead would have a dual effect on the spectrum. First, it would produce a similar effect to an abundance gradient by stretching the spectral lines. Second, it would enhance the abundance gradient in the case of chemical equilibrium, further impacting the observed spectral features. Incorporating more realistic temperature and abundance gradients in future retrieval analyzes will be essential to better constrain the abundances of ionized species.}

\section{Conclusions}

This study presents an in-depth analysis of TOI-1518\,b, an ultra-hot Jupiter. It uses transit observations, from MAROON-X on Gemini-N, to explore its atmospheric dynamics and chemical composition. Our findings offer new insights into the unique characteristics of this extreme exoplanet. {Using the SOPHIE spectrograph, we also report the first significant detection of TOI-1518\,b in radial velocity. This allows us to better characterize the system, and in particular to measure the planetary mass $M_P = 1.83 \pm 0.47$~M$_{\rm{Jup}}$.}

Our cross-correlation analysis focused on detecting atomic and molecular species within the atmosphere. We report the detection of 14 different species. High-resolution spectroscopy allowed us to identify ionized and neutral species with ionized metals such as Fe+, Ca+, and Ti+. The detection of these ionized species, as opposed to their neutral counterparts, underscores the extreme temperatures of TOI-1518\,b's atmosphere, where thermal ionization is significant. Additionally, we reported the detection of vanadium oxide (VO), a critical absorber that has implications for thermal inversions in ultra-hot Jupiters. The presence of VO adds a crucial piece to the puzzle of understanding the thermal structure and chemical processes in such extreme environments. This was possible thanks to the newer HyVO line list that should be used in other studies were detection of VO failed with previous line lists.

We investigated the atmospheric wind dynamics by analyzing the blueshift of multiple species in the observed spectra as a function of the planet's orbital phase. The blueshift of iron is consistent with previous observations of other ultra-hot Jupiters. By combining the signal from different species, particularly Fe+ and Fe, and comparing them with GCMs , we conclude that a strong drag is needed ($\tau_{\rm{drag}}=10^{3}--10^{4}s$ to explain the data. {Fe+ seem to need more substantial drag compared to Fe, which can be due to the increased Ohmic drag in the planetary dayside. Finally, we find that the trail of Ca+ is very different from the other species, which is likely due to Ca+ probing the escaping atmosphere above the Roche lobe while all other species probe deeper atmospheric pressures.}

{ The retrieval analysis provided constraints on the abundance of various chemical species in the atmosphere. We report an abundance of iron of $ {\log_{10}}$Fe$=-4.88^{+0.63}_{-0.76}$, which is consistent with a solar enrichment.  The retrieval being driven mainly by iron, we believe that the iron abundances are robust. However, most other species are detected at a different $K_{\rm{p}}$/$V_{\rm{res}}$, which our retrieval framework does not consider. As a consequence, we believe that the retrieved abundance for the other species are strongly biased. We study in more detail this bias for the specific case of VO, for which our main retrieval does not find a signal, but where a retrieval without iron lead to a detection at a different temperature, $K_{\rm{p}}$, $V_{\rm{res}}$ and an abundance of 1.10 to 45708.81 times solar.}

Overall, our study demonstrates the power of combining high-resolution spectroscopy with advanced modeling techniques to probe the atmospheres of ultra-hot Jupiters. They shed light on the specific properties of TOI-1518\,b and contribute to the broader understanding of atmospheric dynamics and chemistry in these extreme exoplanets. Future studies should continue to refine these models and expand observational efforts to explore the diversity and complexity of ultra-hot Jupiter atmospheres.

\begin{acknowledgements}
    This work was partially funded by the French National Research Agency (ANR) project EXOWINDS (ANR-23-CE31-0001-01).\\
    This work was supported by the French government through the France 2030 investment plan managed by the National Research Agency (ANR), as part of the Initiative of Excellence Université Côte d’Azur under reference number ANR-15-IDEX-01. The authors are grateful to the Université Côte d’Azur’s Center for High-Performance Computing (OPAL infrastructure) for providing resources and support.\\
    J.P.W. acknowledges support from the Trottier Family Foundation via the Trottier Postdoctoral Fellowship.\\
    M.R.L.\ and J.L.B.\ acknowledge support from NASA XRP grant 80NSSC19K0293 and NSF grant AST-2307177. J.L.B.\ acknowledges funding for the MAROON-X project from the David and Lucile Packard Foundation, the Heising-Simons Foundation, the Gordon and Betty Moore Foundation, the Gemini Observatory, the NSF (award number 2108465), and NASA (grant number 80NSSC22K0117).\\
    This work uses observations secured with the SOPHIE spectrograph at the 1.93-m telescope of Observatoire Haute-Provence, France, with the support of its staff. \\
    This work was  supported by the ``Programme National de Plan\'etologie'' (PNP) of CNRS/INSU, 
    and CNES.
\end{acknowledgements}

%choix du style de la biblio

\bibliographystyle{aa}

%inclusion de la biblio

\bibliography{aanda.bib}

\begin{appendix}

\section{Measurement of the planetary mass}\label{Sophie}

{\citealt{2021AJ....162..218C} obtained a hint (less than 2$\sigma$ significance) for a radial-velocity  (RV) detection of TO-1518b using the FIES spectrograph. They reported an RV semi-amplitude of  $152 \pm 75$~m/s, which they converted to a 2-$\sigma$ upper limit of 281~m/s, corresponding to a planetary upper mass limit of 2.3~M$_{\rm{Jup}}$. }

{With the goal to improve the constraint on the planetary mass, we observed TOI-1518 with 
the SOPHIE spectrograph at the 1.93-m telescope of the Observatoire de Haute-Provence, France.
SOPHIE is a stabilized \'echelle spectrograph dedicated to high-precision RV measurements 
\citep{2008SPIE.7014E..0JP,2009A&A...505..853B,2013A&A...549A..49B}. 
We used its high-resolution mode (resolving power $R=75\,000$) and fast 
readout mode. We obtained exposures at 14 different epochs between December 2023 and January 2024.}
{Exposure times ranged between 3.9 and 16.8 minutes depending on weather conditions, allowing 
signal-to-noise from 35.6 to 47.3 to be reached per pixel at 550\,nm. The log of the observations
 is reported in Table~\ref{table_rv}.}

{The RVs were extracted with the SOPHIE pipeline, as presented by \citet{2009A&A...505..853B} and refined by \citet{2024A&A...681A..55H}. It derives cross 
correlation functions (CCF) from standard numerical masks corresponding to different spectral types. }
{As expected from the known rapid rotation speed of the star, the derived CCFs were broad, 
with typical FWHM around 100~km/s. So to measure the RV, instead of fitting a Gaussian profile 
as it is standardly done for slow rotators, here we fitted the CCF with a profile constructed from the convolution of a Gaussian and a rotational profile following \citet{2022oasp.book.....G}. This provided good fits of the CCFs.
The Moonlight pollution, estimated using the second SOPHIE fiber aperture that is targeted on the sky, 2' away from the first one pointing toward the star, was negligible.} 

{The FWHM of the Gaussian is unknown, but is  expected to be small by comparison to the 
FWHM of the rotational profile. We attempted free and fixed values for that Gaussian FWHM, 
which did not significantly modify our results. We finally fixed the FWHM of the Gaussian profile to 7~km/s on each exposure, which is a common value for such kind of stars.
The FWHM of the rotational  {profile} was free to vary among exposures, and we obtained values 
corresponding to $v \sin i_* = 78.9$~km/s, with a dispersion of $\pm 1.1$~km/s. This is in good 
agreement with the values $v \sin i_* = 85 \pm 6$~km/s reported by \citealt{2021AJ....162..218C}.
The measured RVs (see Table~\ref{table_rv}) are derived from the fitted center of the Gaussian and rotational profiles, constrained to be identical for a given epoch.}

{We phase-folded those SOPHIE's RVs  using the 1.9-d period derived from the transits observed by TESS. 
This provided a significant RV semi-amplitude $K=232^{+65}_{-64}$~m/s, in phase with 
the transit epoch measured with TESS. This agrees with the upper limit reported by 
\citet{2021AJ....162..218C}. So we concluded our SOPHIE data allow the planet
TOI-1518\,b to be significantly detected, for the first time with RVs.}

{To refine the system parameters, we made a joined fit of the TESS transit light curves and the available RVs. 
Following \citet{2025A&A...694A..36H}, we fitted that dataset using the EXOFASTv2 package \citep{2013PASP..125...83E,2017ascl.soft10003E,2019arXiv190709480E}.
\citet{2021AJ....162..218C} used TESS sectors 17 and 18 covering TOI-1518, observed in FFI mode so with a long-cadence 30-min sampling, in addition to five ground-based photometric follow up. 
Here, we used both those TESS sectors, as well as the three additional TESS sectors covering TOI-1518  now available, which have a short-cadence 2-min sampling (sectors 57, 58, and 77). So by comparison to \citet{2021AJ....162..218C}, the number of TOI-1518\,b individual transits included in our fit increases from 24 to 56. And in addition to our new SOPHIE RVs, we also used the FIES RVs published by \citet{2021AJ....162..218C}. }

{\citet{2021AJ....162..218C} reported a negligible eccentricity 
($e = 0.0031^{+0.0047}_{-0.0022}$). We tested eccentric and circular orbits
without detecting significant differences, and finally 
adopted here a circular orbit.  According to \citet{2024PASJ..tmp...34W}, 
the system shows a change in the impact parameter $b$ of the planet as a function of time. 
Over the 4.5-yr span of the TESS observations we use, the expected effect is of the order of 0.05 on $b$. This is not taken into account here, and we report averaged values.
This has no significant impact on the reported planetary mass.}

\begin{figure*}[ht]
 \centering
\includegraphics[width=1.9\columnwidth]{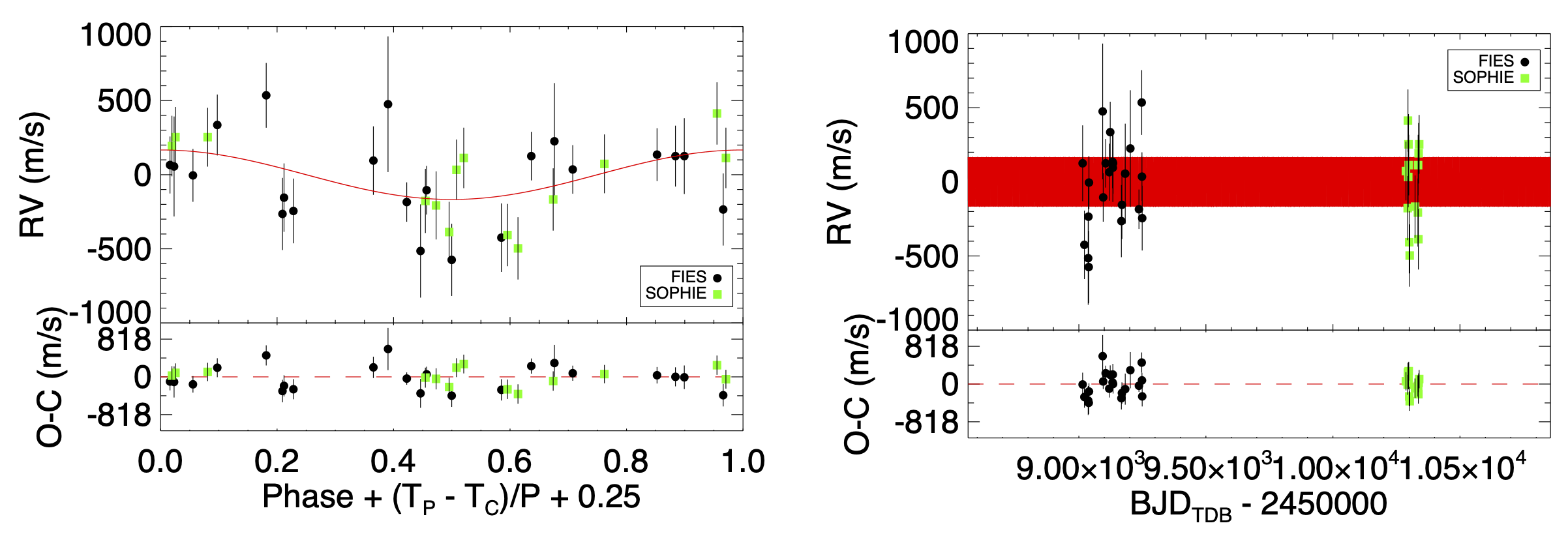}
  \caption{Radial velocities of TOI-1518 as a function of the planetary orbital phase (left) and time (right). 
  SOPHIE (Table~\ref{table_rv}) and FIES \citep{2021AJ....162..218C} measurements are plotted in 
  green and black, respectively, together with their 1-$\sigma$ error bars. The fitted Keplerian orbit 
  (Table~\ref{table_param_fit}) 
  is overplotted in~red. The bottom panels show the residuals.}
  \label{fig_RVs}
\end{figure*}

\begin{figure*}[ht]
 \centering
\includegraphics[width=1\columnwidth]{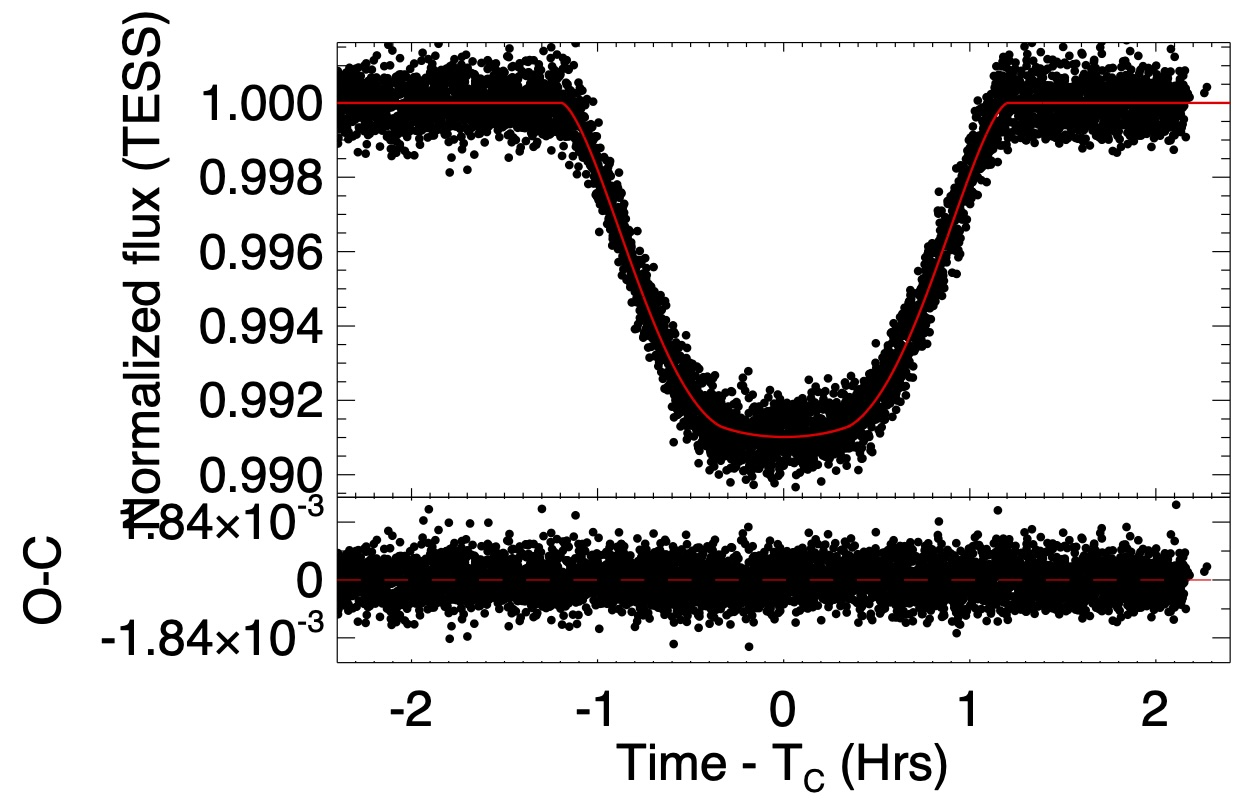}

  \caption{Phase-folded TESS transit light curve of TOI-1518. The fitted 
  Keplerian orbit (Table~\ref{table_param_fit}) is overplotted in red. The bottom panel shows the~residuals.}
  \label{fig_TESS}
\end{figure*}

{Our fitted RVs and transit light curve are shown in Fig.~\ref{fig_RVs} and \ref{fig_TESS}, respectively. We finally measured an RV semi-amplitude $K = 192 ^{+48}_{-49}$~m/s, corresponding to a planetary mass $1.83 \pm 0.47$~M$_{\rm{Jup}}$. 
 The full results of our fit are reported in Table~\ref{table_param_fit}.
Our derived parameters agree with those presented by \citealt{2021AJ....162..218C}, 
with improved uncertainties for most of~them.}

\begin{table}[t]
\caption{SOPHIE measurements of the planet-host star TOI-1518}
\begin{center}
\begin{tabular}{ccccc}
\toprule
BJD$_{\rm{UTC}}$ & RV & $\pm$$1\,\sigma$ & exp. & SNR per pix \\
-2\,460\,000 & (km/s) & (km/s) & (sec) &  at 550\,nm \\
\midrule
287.28150 & -11.64 & 0.11 &  1010 & 46.3  \\   
294.30716 & -11.89 & 0.14 &  543  & 46.7  \\   
295.25968 & -11.30 & 0.13 &  284  & 46.6  \\   
296.31131 & -11.68 & 0.12 &  268  & 46.9  \\   
297.29585 & -11.46 & 0.12 &  271  & 46.6  \\   
298.23770 & -11.60 & 0.12 &  233  & 46.5  \\   
300.28312 & -12.12 & 0.13 &  803  & 41.6  \\   
302.22014 & -12.21 & 0.13 &  662  & 46.4  \\   
323.26383 & -11.88 & 0.13 &  452  & 47.1  \\   
334.29677 & -11.92 & 0.16 &  904  & 35.6  \\   
335.24376 & -11.60 & 0.12 &  593  & 46.6  \\   
336.24214 & -12.10 & 0.12 &  530  & 46.9  \\   
337.23923 & -11.52 & 0.12 &  237  & 46.7  \\   
339.25846 & -11.46 & 0.11 &  407  & 47.3  \\   
\bottomrule
\end{tabular}
\end{center}
\label{table_rv}
\end{table}

\begin{table*}[t]
\caption{Median values and 68\% confidence interval for the TOI-1518 system.
See Table 3 in \citet{2019arXiv190709480E} for a detailed description of all parameters.}
\begin{center}
\begin{tabular}{lccc}
\toprule
Parameter & Units & \multicolumn{2}{c}{Values} \\
\midrule
\smallskip\\\multicolumn{2}{l}{Stellar Parameters:}&TOI-1518\smallskip\\
~~~~$M_*$\dotfill &Mass (M$_{\odot}$)\dotfill &$1.88^{+0.13}_{-0.12}$\\
~~~~$R_*$\dotfill &Radius (R$_{\odot}$)\dotfill &$1.942\pm0.043$\\
~~~~$L_*$\dotfill &Luminosity (L$_{\odot}$)\dotfill &$9.64^{+0.69}_{-0.65}$\\
~~~~$\rho_*$\dotfill &Density (cgs)\dotfill &$0.3622^{+0.0055}_{-0.0054}$\\
~~~~$\log{g}$\dotfill &Surface gravity (cgs)\dotfill &$4.136\pm0.011$\\
~~~~$T_{\rm eff}$\dotfill &Effective Temperature (K)\dotfill &$7299^{+98}_{-100}$\\
~~~~$[{\rm Fe/H}]$\dotfill &Metallicity (dex)\dotfill &$-0.10\pm0.12$\\
\smallskip\\\multicolumn{2}{l}{Planetary Parameters:}&TOI-1518b\smallskip\\
~~~~$P$\dotfill &Period (days)\dotfill &$1.90261131\pm0.00000043$\\
~~~~$R_P$\dotfill &Radius (R$_{\rm Jup}$)\dotfill &$1.878\pm0.042$\\
~~~~$M_P$\dotfill &Mass (M$_{\rm Jup}$)\dotfill &$1.83\pm0.47$\\
~~~~$T_T$\dotfill &Time of transit (\ensuremath{\rm {BJD_{TDB}}})\dotfill &$2458787.04943\pm0.00028$\\
~~~~$T_0$\dotfill &Optimal conjunction Time (\ensuremath{\rm {BJD_{TDB}}})\dotfill &$2459983.791942\pm0.000066$\\
~~~~$a$\dotfill &Semi-major axis (AU)\dotfill &$0.03712\pm0.00082$\\
~~~~$i$\dotfill &Inclination (degrees)\dotfill &$77.626\pm0.097$\\
~~~~$T_{eq}$\dotfill &Equilibrium temperature (K)\dotfill &$2546^{+35}_{-36}$\\
~~~~$\tau_{\rm circ}$\dotfill &Tidal circularization timescale (Gyr)\dotfill &$0.0059\pm0.0015$\\
~~~~$K$\dotfill &RV semi-amplitude (m/s)\dotfill &$192^{+48}_{-49}$\\
~~~~$R_P/R_*$\dotfill &Radius of planet in stellar radii \dotfill &$0.09939^{+0.00050}_{-0.00051}$\\
~~~~$a/R_*$\dotfill &Semi-major axis in stellar radii \dotfill &$4.109^{+0.021}_{-0.020}$\\
~~~~$\delta$\dotfill &Transit depth (fraction)\dotfill &$0.00988\pm0.00010$\\
~~~~$Depth$\dotfill &Flux decrement at mid transit \dotfill &$0.00988\pm0.00010$\\
~~~~$\tau$\dotfill &Ingress/egress transit duration (days)\dotfill &$0.03562^{+0.00071}_{-0.00069}$\\
~~~~$T_{14}$\dotfill &Total transit duration (days)\dotfill &$0.09978^{+0.00034}_{-0.00033}$\\
~~~~$T_{FWHM}$\dotfill &FWHM transit duration (days)\dotfill &$0.06415^{+0.00081}_{-0.00079}$\\
~~~~$b$\dotfill &Transit Impact parameter \dotfill &$0.8806^{+0.0025}_{-0.0027}$\\
~~~~$\delta_{S,2.5\mu m}$\dotfill &Blackbody eclipse depth at 2.5$\mu$m (ppm)\dotfill &$1378\pm24$\\
~~~~$\delta_{S,5.0\mu m}$\dotfill &Blackbody eclipse depth at 5.0$\mu$m (ppm)\dotfill &$2271^{+26}_{-25}$\\
~~~~$\delta_{S,7.5\mu m}$\dotfill &Blackbody eclipse depth at 7.5$\mu$m (ppm)\dotfill &$2633^{+27}_{-26}$\\
~~~~$\rho_P$\dotfill &Density (cgs)\dotfill &$0.343^{+0.087}_{-0.088}$\\
~~~~$logg_P$\dotfill &Surface gravity \dotfill &$3.110^{+0.097}_{-0.13}$\\
~~~~$\Theta$\dotfill &Safronov Number \dotfill &$0.0385^{+0.0097}_{-0.0098}$\\
~~~~$\langle F \rangle$\dotfill &Incident Flux (10$^9$ erg s$^{-1}$ cm$^{-2}$)\dotfill &$9.54^{+0.53}_{-0.52}$\\
~~~~$T_S$\dotfill &Time of eclipse (\ensuremath{\rm {BJD_{TDB}}})\dotfill &$2458788.00074\pm0.00028$\\
~~~~$T_A$\dotfill &Time of Ascending Node (\ensuremath{\rm {BJD_{TDB}}})\dotfill &$2458788.47639\pm0.00028$\\
~~~~$T_D$\dotfill &Time of Descending Node (\ensuremath{\rm {BJD_{TDB}}})\dotfill &$2458787.52509\pm0.00028$\\
~~~~$M_P/M_*$\dotfill &Mass ratio \dotfill &$0.00093\pm0.00024$\\
~~~~$d/R_*$\dotfill &Separation at mid transit \dotfill &$4.109^{+0.021}_{-0.020}$\\
~~~~$P_T$\dotfill &A priori non-grazing transit prob \dotfill &$0.2192^{+0.0012}_{-0.0011}$\\
~~~~$P_{T,G}$\dotfill &A priori transit prob \dotfill &$0.2676\pm0.0013$\\
\smallskip\\\multicolumn{2}{l}{Wavelength Parameters:}&TESS\smallskip\\
~~~~$u_{1}$\dotfill &linear limb-darkening coeff \dotfill &$0.165^{+0.035}_{-0.034}$\\
~~~~$u_{2}$\dotfill &quadratic limb-darkening coeff \dotfill &$0.326\pm0.035$\\
\smallskip\\\multicolumn{2}{l}{Telescope Parameters:}&FIES&SOPHIE\smallskip\\
~~~~$\gamma_{\rm rel}$\dotfill &Relative RV Offset (m/s)\dotfill &$-14787^{+52}_{-53}$&$-11708^{+57}_{-58}$\\
~~~~$\sigma_J$\dotfill &RV Jitter (m/s)\dotfill &$0.00^{+110}_{-0.00}$&$172^{+66}_{-53}$\\
~~~~$\sigma_J^2$\dotfill &RV Jitter Variance \dotfill &$-16000^{+29000}_{-17000}$&$30000^{+27000}_{-16000}$\\
\smallskip\\\multicolumn{2}{l}{Transit Parameters:}&TESS long cadence &TESS short cadence\smallskip\\
~~~~$\sigma^{2}$\dotfill &Added Variance \dotfill &$0.000000234^{+0.000000031}_{-0.000000026}$&$0.0000000225^{+0.0000000060}_{-0.0000000059}$\\
~~~~$F_0$\dotfill &Baseline flux \dotfill &$1.0017\pm0.0044$&$1.0004^{+0.0018}_{-0.0014}$\\
\end{tabular}
\end{center}
\label{table_param_fit}
\end{table*}

\FloatBarrier 

\onecolumn
\section{Cross-correlation-function}

\begin{figure}[h!]
   \centering
   \includegraphics[width=6cm]{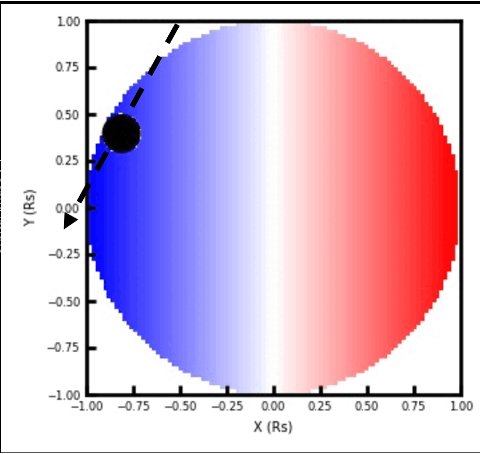}
   \caption{Qualitative geometry of the transit of TOI-1518\,b. This planet is highly misaligned with the star (impact parameter = 0.9).}
    \label{FigGeometry}%
\end{figure}
\FloatBarrier 

\begin{figure*}[h!]
   \centering
   \includegraphics[width=15cm]{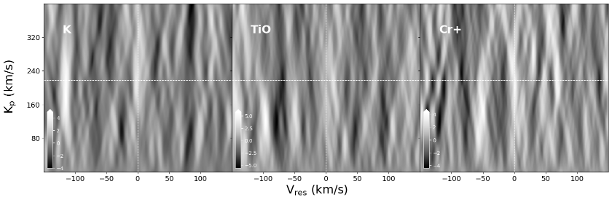}
   \caption{\textbf{All K$_{p}-V_{res}$ diagram for non-detected species in TOI-1518\,b dataset}. The white cross indicates the expected location of the planetary signal, which assumes a static atmosphere.  
      The signal observed at K$_{p}$ around 100 km.s$^-1$ and V$_{res}$ around -60 km.s$^-1$ in some diagrams is an artifact due to the Doppler shadow. 
             }
    \label{Fig_no_detection}%
\end{figure*}

\begin{figure}[!thbp]
   \centering
   \includegraphics[width=15cm]{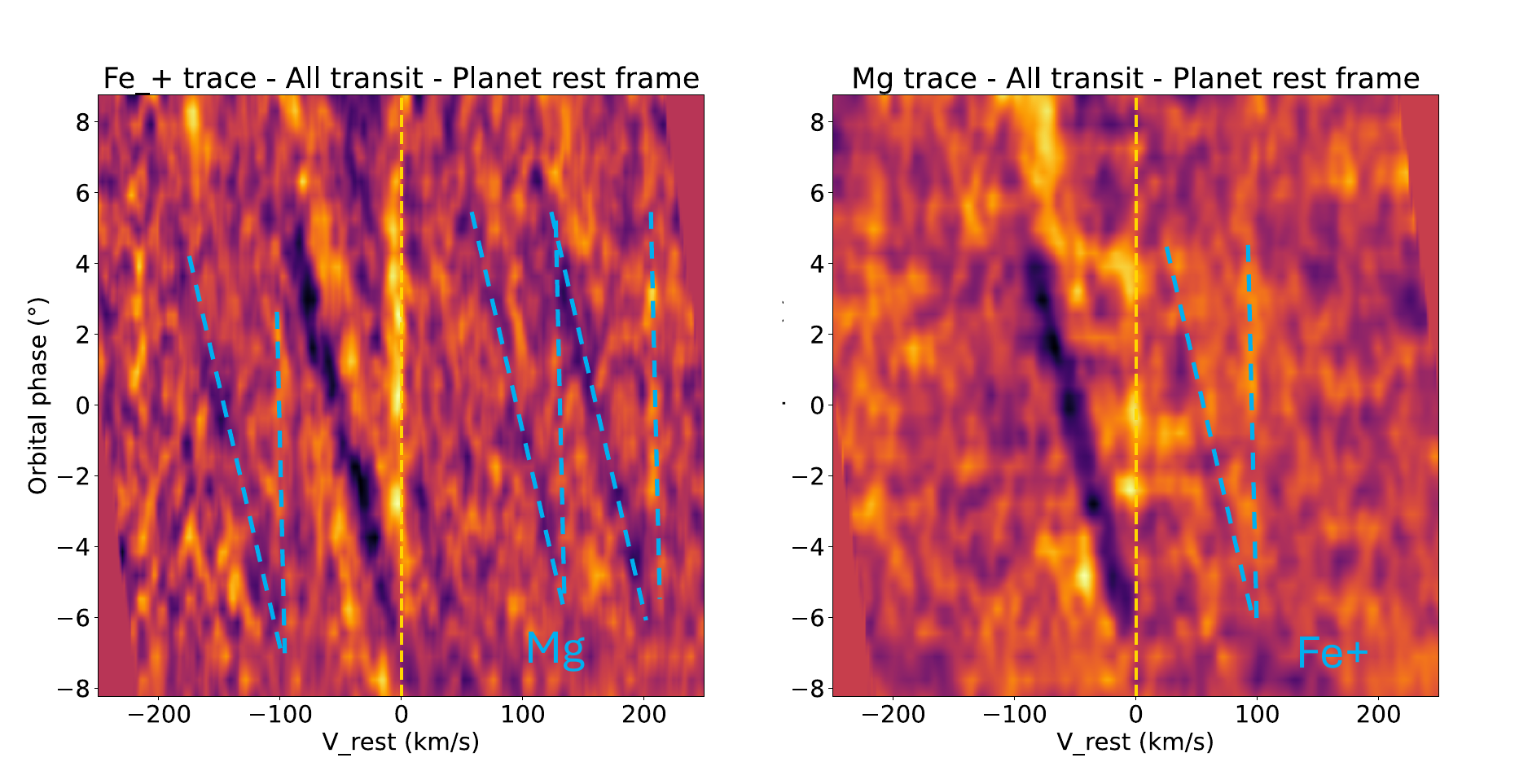}
   \caption{Trails of Fe + and Mg with contamination from Mg in Fe + trail and Fe + in Mg trail due to the proximity of a strong Fe+ feature in the Magnesium triplet.}
    \label{Fig_Fe_Mg}%
\end{figure}

\begin{figure}[!thbp]
   \centering
      \includegraphics[width=14cm]{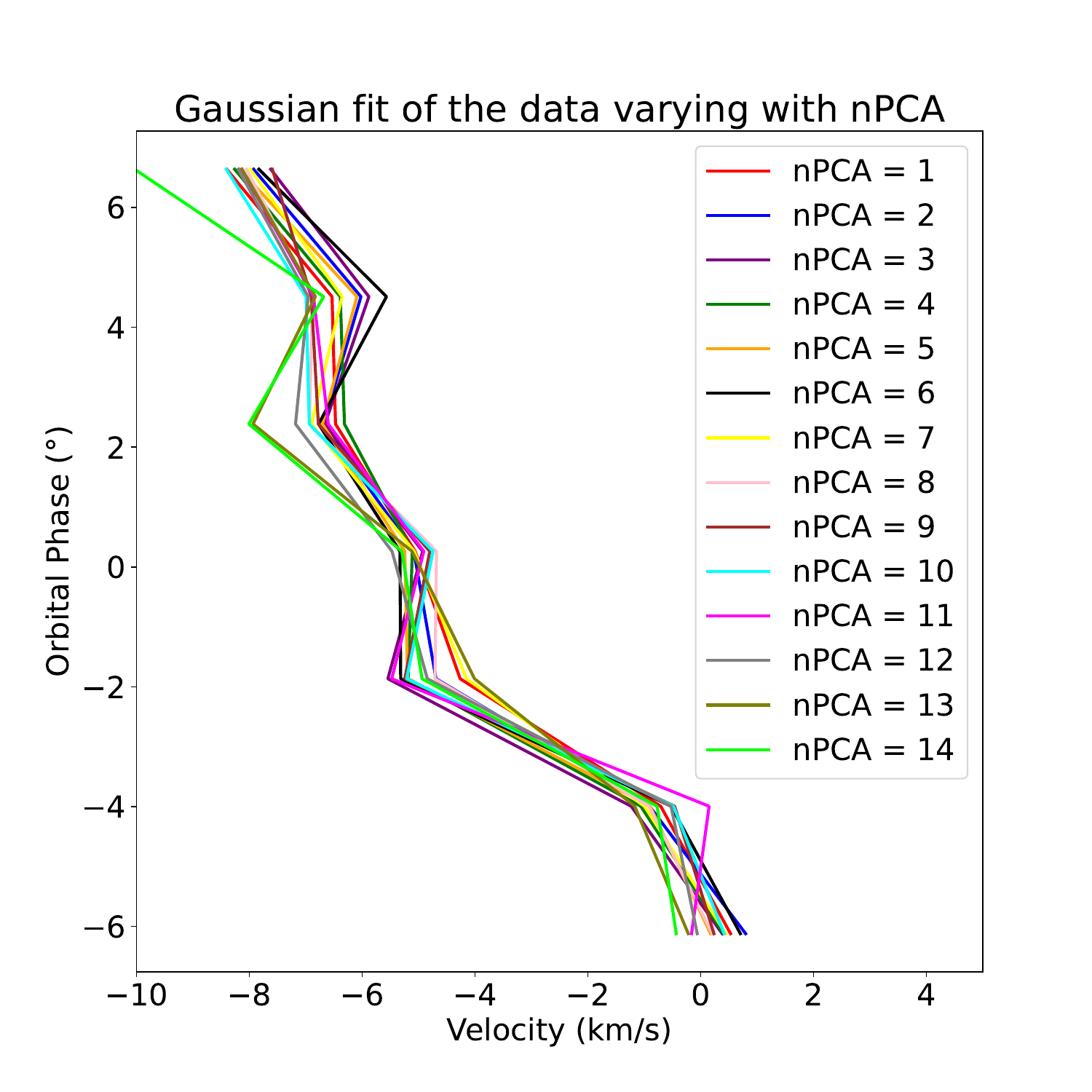}\\
      \caption{Measured Doppler shifts for Fe as a function of orbital phase angle. Different colors represent different numbers of components removed from the data.}
         \label{Fig_nPCA}
\end{figure}

\begin{figure}[!thbp]
   \centering
   \includegraphics[width=14cm]{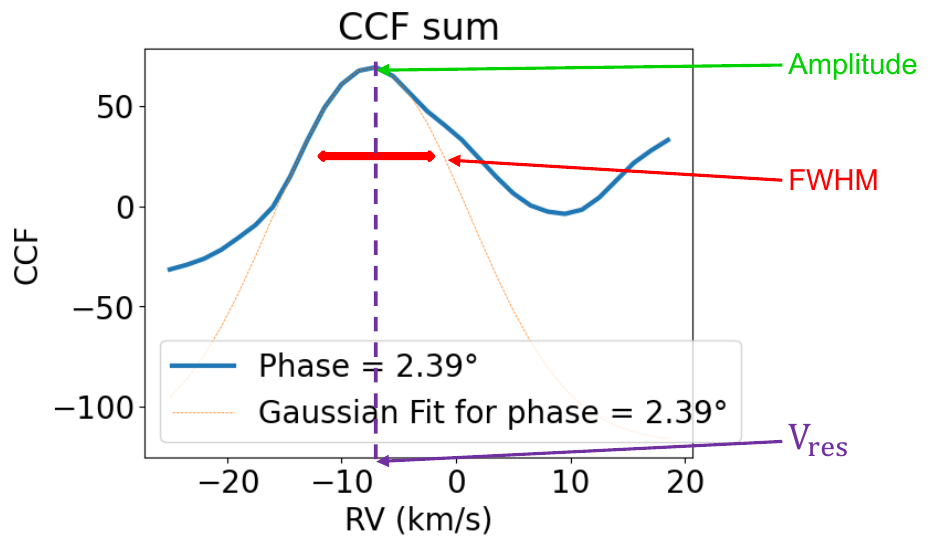}
   \caption{Example of the Gaussian fit applies to one binned CCF result. The blue line is the CCF binned for one orbital phase. In the yellow dashed line, the Gaussian fit performs where we infer the three parameters presented in Fig.\ref{Fig_trails_param} and Fig.\ref{Fig:Gauss_model}.}
    \label{Fig_Gauss_fit}%
\end{figure}

\begin{figure*}[!thbp]
   \centering
      \includegraphics[width=17cm]{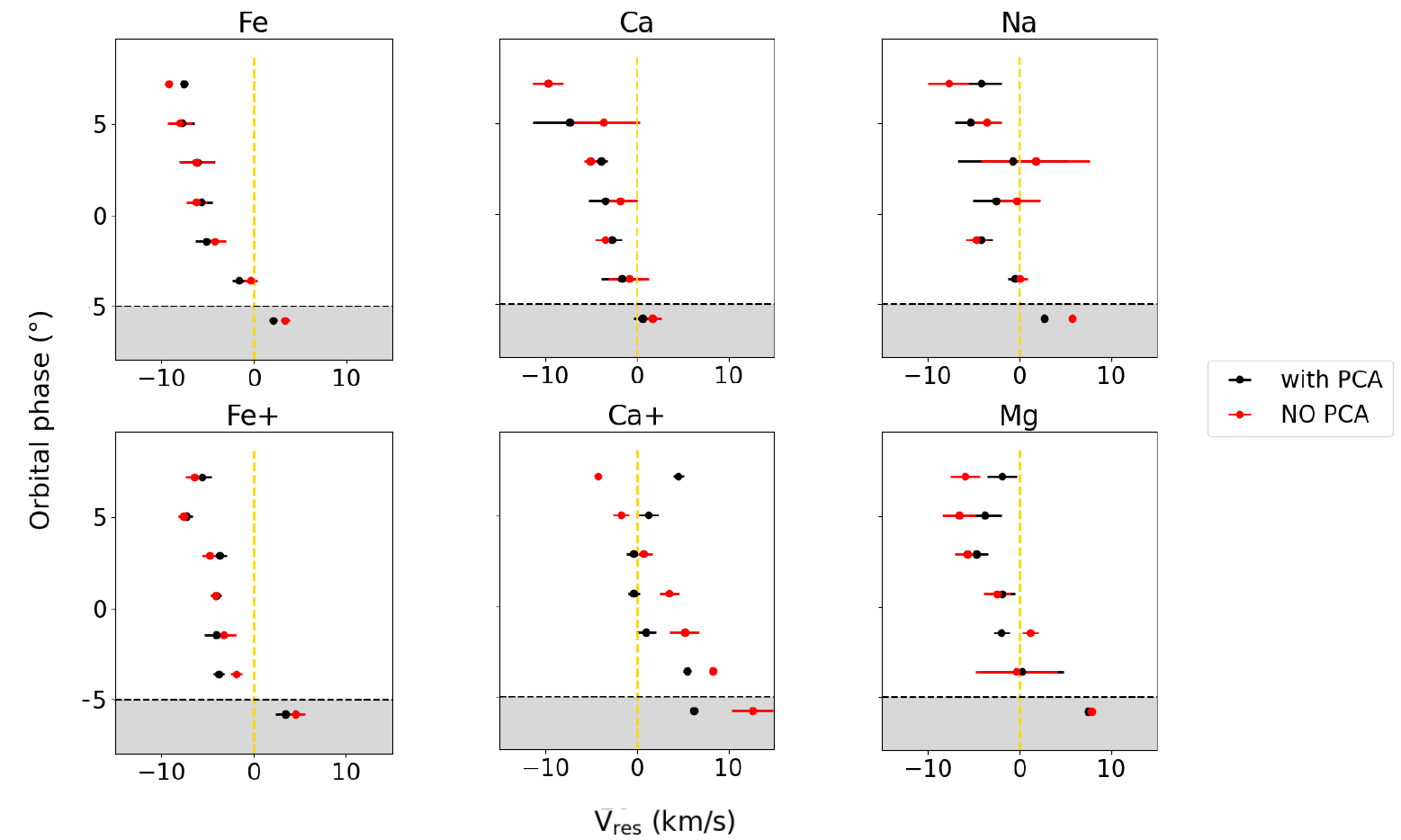}\\
      \caption{Same as lower panel of Fig.\ref{Diff_Trail} but with the results of the Gaussian fit for the study with and without PCA.}
         \label{Fig_comp_trails}
\end{figure*}

\begin{figure*}[!htbp]
   \centering
      \includegraphics[width=18cm]{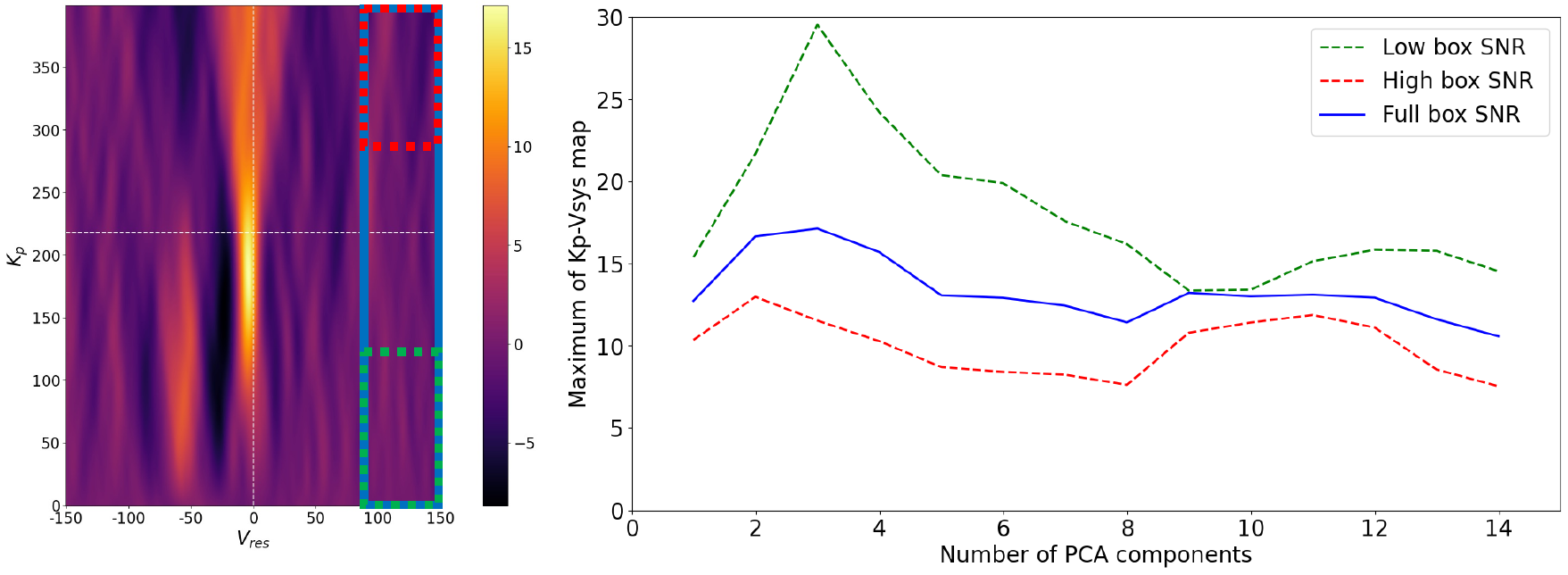}
      \caption{Comparison of the S/N level obtained as function of number of PCA components with different boxes used for the calculation of the standard deviation of the Fe $K_p$-$V_{\rm res}$ map.}
         \label{Diff_comp}
\end{figure*}
\FloatBarrier 

\clearpage
\section{Global circulation models }

\begin{figure*}[!thbp]
   \centering
      \includegraphics[width=18cm]{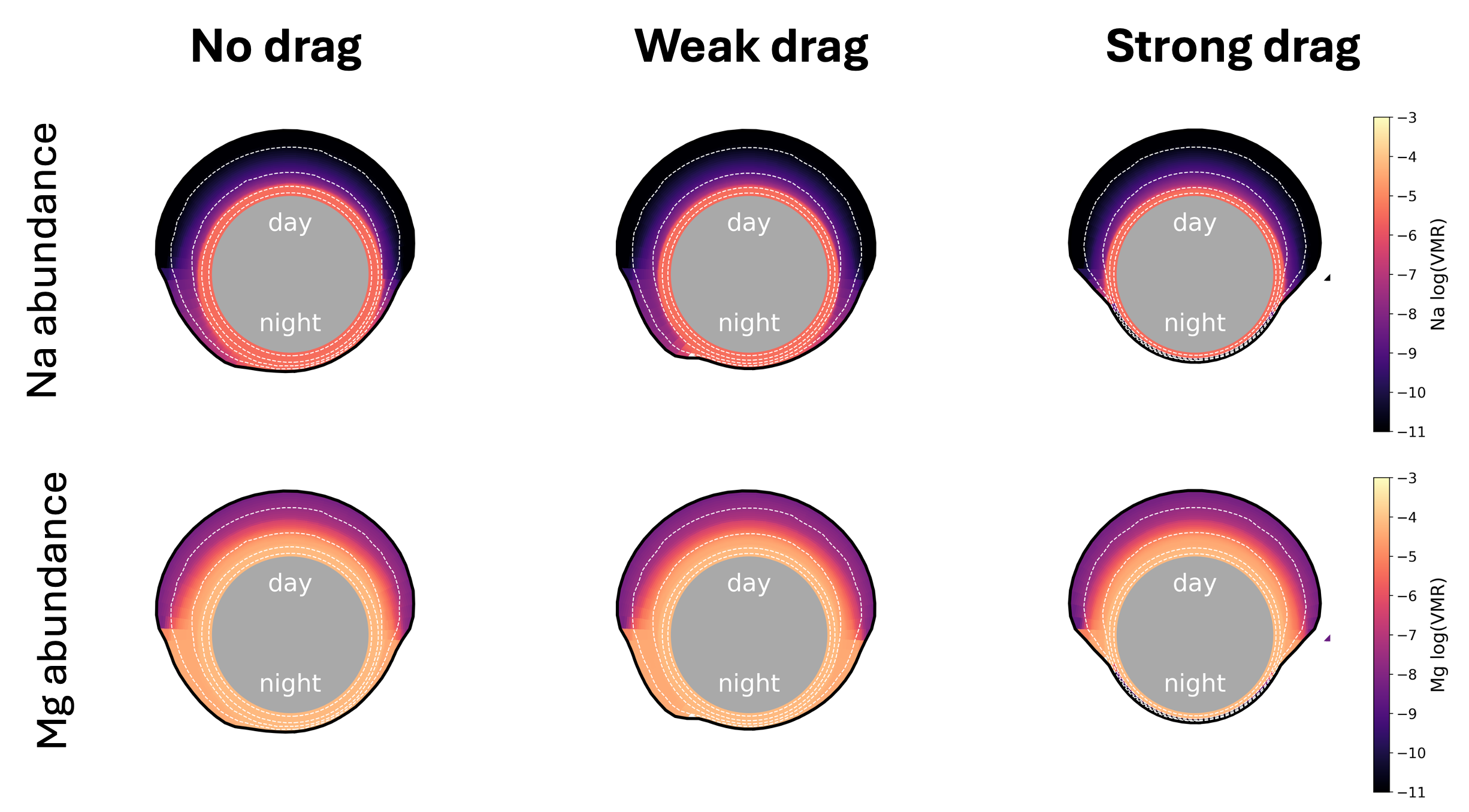}
      \caption{Same as figure \ref{fig: GCMs} with sodium and magnesium abundances.
              }
         \label{fig: GCMs_Na_Mg}
\end{figure*}
\FloatBarrier 

\clearpage
\section{Retrieval analysis}

\FloatBarrier 
\begin{figure*}[!thbp]
    \centering
    \includegraphics[width=19cm]{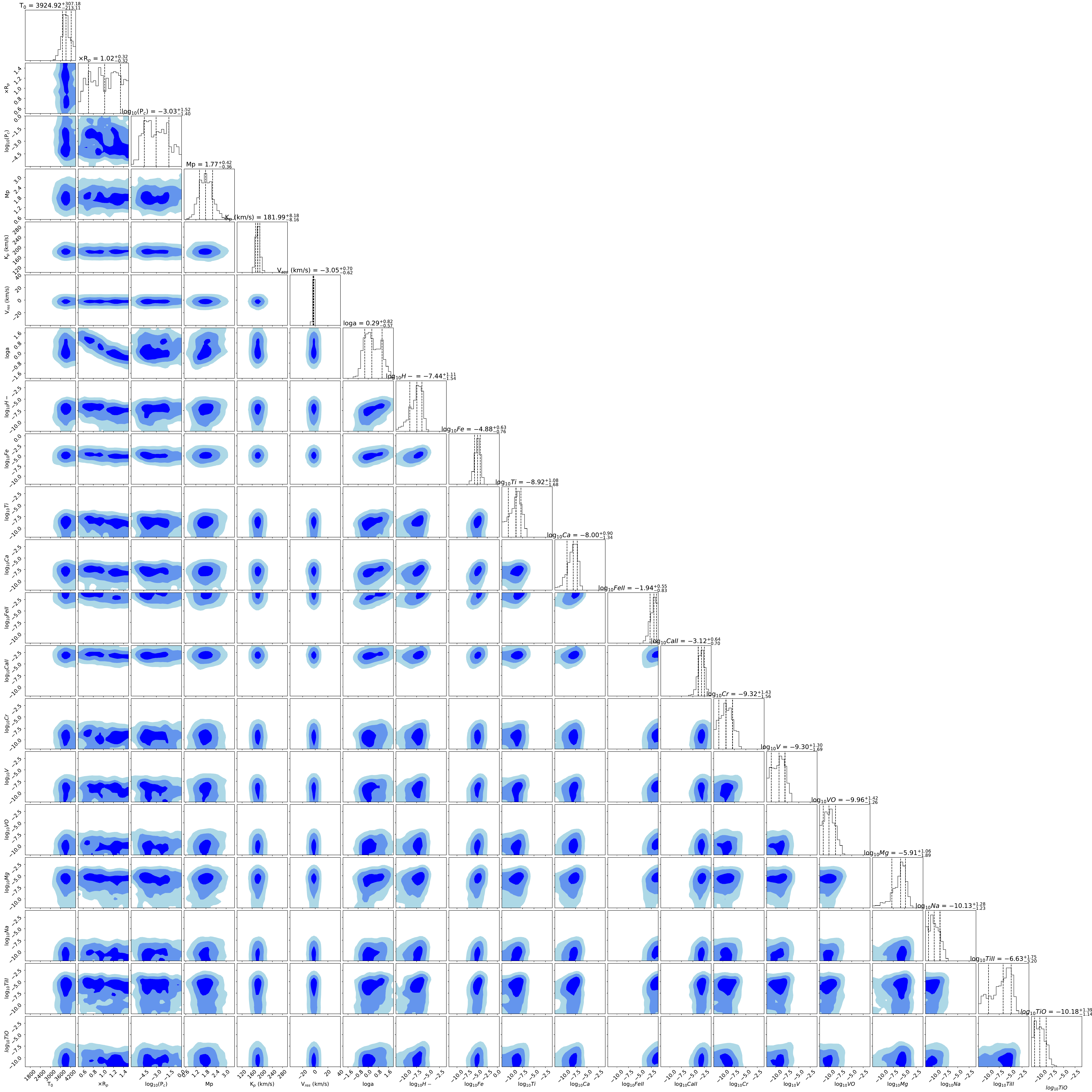}
    \caption{Corner plot of the full retrieval analysis}
    \label{fig:Full_retrieval}
\end{figure*}

\begin{figure*}[!thbp]
    \centering
    \includegraphics[width=19cm]{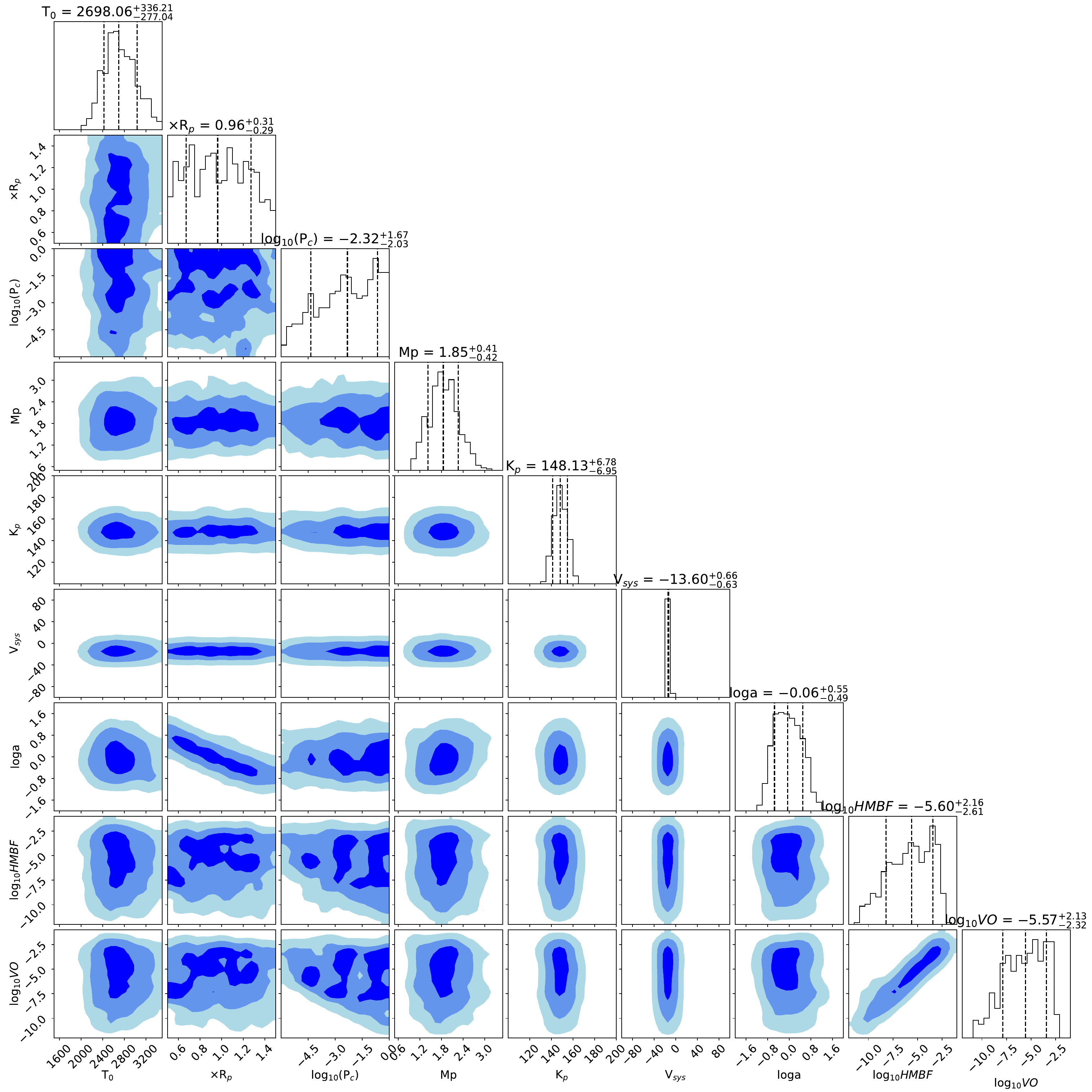}
    \caption{Corner plot of the single species retrieval with only VO}
    \label{fig:Retriev_Vo_only}
\end{figure*}

\end{appendix}

 \clearpage
%--------------------------------------------------------------------

%\begin{thebibliography}{}

%\end{thebibliography}

\end{document}